\documentclass[12pt]{article}
\usepackage{color}

\usepackage{graphicx}
\usepackage{url}
\usepackage{caption}
\usepackage{subcaption}
\usepackage{float}
\usepackage{array}

\usepackage{epsfig}
\usepackage{times}
\usepackage{amsmath}

\usepackage[round]{natbib}
\usepackage{appendix}
\usepackage{amssymb}
\usepackage{multirow}
\usepackage{lineno}
\usepackage{setspace}
\usepackage{todonotes}
\usepackage{booktabs}

\usepackage{authblk}

\definecolor{darkblue}{RGB}{0,0,147}
\usepackage[final]{hyperref}
\hypersetup{
	colorlinks=true, 
	linktocpage=true,
	breaklinks=true, 
	pageanchor=true,
	plainpages=false, 
	hypertexnames=true,
	urlcolor=darkblue, 
	linkcolor=darkblue,
	citecolor=darkblue, 
	pdftitle={SE-KGE: A Location-Aware Knowledge Graph Embedding Model for Geographic Question Answering and Spatial Semantic Lifting},
	pdfauthor={Gengchen Mai, Krzysztof Janowicz, Ling Cai, Rui Zhu, Blake Regalia, Bo Yan, Meilin Shi, Ni Lao}
	pdfsubject={Geographic Query Answering},
	pdfkeywords={Knowledge Graph Embedding, Location Encoding, Spatially Explicit Model, Geographic Question Answering, Spatial Semantic Lifting},
	pdfcreator={pdfLaTeX},
	pdfproducer={LaTeX via TeX Live on GNU/LINUX}
}

\usepackage [autostyle, english = american]{csquotes}
\MakeOuterQuote{"}

\usepackage{listings}
\usepackage{authblk}
\usepackage{changes}
\usepackage{adjustbox}
\usepackage{rotating}

\let\oldhat\hat
\renewcommand{\vec}[1]{\mathbf{#1}}
\renewcommand{\hat}[1]{\oldhat{\mathbf{#1}}}

\title{
\spexkge: 
A Location-Aware Knowledge Graph Embedding Model
for Geographic Question Answering and Spatial Semantic Lifting
}

\author[1]{Gengchen Mai}
\author[1]{Krzysztof Janowicz}
\author[1]{Ling Cai}
\author[1]{Rui Zhu}
\author[1]{Blake Regalia}
\author[2]{Bo Yan}
\author[1]{Meilin Shi}
\author[3]{Ni Lao}
\affil[1]{STKO Lab, University of California, Santa Barbara, CA, USA, 93106}
\affil[ ]{\textit {\{gengchen\_mai, janowicz, lingcai, ruizhu, regalia, meilinshi\}@ucsb.edu}}
\affil[2]{LinkedIn Corporation, Mountain View, CA, USA, 94043}
\affil[ ]{\textit {boyan1@linkedin.com}}
\affil[2]{SayMosaic Inc., Palo Alto, CA, USA, 94303}
\affil[ ]{\textit {ni.lao@mosaix.ai}}

\date{}
\usepackage{amsthm}
\usepackage{amsmath}
\usepackage{mathabx}
\usepackage{xspace}
\usepackage{lscape}
\usepackage{amsthm}
\newtheorem{problem}{Task}
\newtheorem{definition}{Definition}
\theoremstyle{definition}

\DeclareMathOperator*{\argmax}{arg\,max}

\def\bx{\vec{x}}

\def\blockdiag{diag}

\newcommand{\queryA}{q_{A}}
\newcommand{\queryB}{q_{B}}
\newcommand{\queryC}{q_{C}}

\newcommand{\qagoldfun}{\varphi}
\newcommand{\kgqefun}{\mathbf{\Phi}}
\newcommand{\spasemliftfun}{\mathbf{\Psi}}

\newcommand{\bgpone}{b_{1}}
\newcommand{\bgptwo}{b_{2}}
\newcommand{\bgpthree}{b_{3}}

\newcommand{\dbgeo}{\textit{DBGeo}}
\newcommand{\spexkge}{\textit{SE-KGE}}
\newcommand{\ssl}{\textit{ssl}}

\newcommand{\pedim}{d^{(x)}}
\newcommand{\fedim}{d^{(\type)}}
\newcommand{\embdim}{d}

\newcommand{\peemb}{\mathbf{\ent}^{(x)}}
\newcommand{\feemb}{\mathbf{\ent}^{(\type)}}

\newcommand{\studyarea}{\mathcal{A}}

\newcommand{\Real}{\mathbb{R}}

\newcommand{\domain}{Domain}
\newcommand{\range}{Range}
\newcommand{\gp}{b}
\newcommand{\ent}{e}
\newcommand{\rel}{r}
\newcommand{\variable}{V}
\newcommand{\typefun}{\Gamma}
\newcommand{\type}{c}
\newcommand{\typeset}{\mathcal{C}}
\newcommand{\embmat}{\mathbf{Z}}
\newcommand{\entemb}{\mathbf{e}}
\newcommand{\varemb}{\mathbf{v}}

\newcommand{\realnum}{\mathbb{R}}
\newcommand{\onehotvec}{\mathbf{h}}
\newcommand{\cgq}{q}

\newcommand{\subject}{h}

\newcommand{\coordsys}{\mathcal{XY}}

\newcommand{\kg}{\mathcal{G}}
\newcommand{\entset}{\mathcal{V}}
\newcommand{\edgeset}{\mathcal{E}}
\newcommand{\relset}{\mathcal{R}}
\newcommand{\triset}{\mathcal{T}}
\newcommand{\triple}{s}
\newcommand{\ptentset}{\entset_{pt}}
\newcommand{\ptfun}{\mathcal{PT}}
\newcommand{\pgonentset}{\entset_{pn}}
\newcommand{\pgonfun}{\mathcal{PN}}
\newcommand{\geofun}{\mathcal{X}}

\newcommand{\kgdefs}{(\entset, \edgeset)}

\newcommand{\gqe}{GQE}
\newcommand{\gqediag}{GQE_{diag}}
\newcommand{\cga}{CGA}

\newcommand{\enc}{Enc}
\newcommand{\featenc}{\enc^{(\type)}}
\newcommand{\spaenc}{\enc^{(x)}}
\newcommand{\encgqe}{\enc^{(\gqe)}}
\newcommand{\enccga}{\enc^{(\cga)}}

\newcommand{\locenc}{LocEnc^{(x)}}
\newcommand{\pemlp}{\mathbf{NN}}

\newcommand{\nscale}{S}

\newcommand{\proj}{\mathcal{P}}
\newcommand{\projmat}{\mathbf{R}}
\newcommand{\projfeatmat}{\projmat_{\rel}^{(\type)}}
\newcommand{\projspamat}{\projmat_{\rel}^{(x)}}
\newcommand{\projspatofeatmat}{\projmat_{\rel}^{(x \type)}}
\newcommand{\projent}{\proj^{(e)}}
\newcommand{\projloc}{\proj^{(x)}}
\newcommand{\projgqe}{\proj^{(\gqe)}}
\newcommand{\projgqediag}{\proj^{(\gqediag)}}

\newcommand{\projcga}{\proj^{(\cga)}}

\newcommand{\inter}{\mathcal{I}}
\newcommand{\intergqe}{\mathcal{I}^{(\gqe)}}
\newcommand{\intercga}{\mathcal{I}^{(\cga)}}

\newcommand{\neisize}{n}

\newcommand{\kgtembed}{\mathbf{H}_{KG}}

\newcommand{\kgtloss}{\mathcal{L}_{KG}}
\newcommand{\qatloss}{\mathcal{L}_{QA}}
\newcommand{\loss}{\mathcal{L}}

\newcommand{\lploss}{\mathcal{L}_{LP}}
\newcommand{\sslloss}{\mathcal{L}_{SSL}}

\newcommand{\cosine}{\mathbf{\Omega}}

\newcommand{\nei}{N}
\newcommand{\neisamp}{N_{\neisize}}
\newcommand{\negsamp}{Neg}

\newcommand{\qasamplesize}{X}

\newcommand{\qaset}{S}

\newcommand{\query}{q}
\newcommand{\answer}{a}

\newcommand{\spasemliftent}{e^{\prime}}
\newcommand{\spasemliftentemb}{\mathbf{e}^{\prime}}

\newcommand{\spasemliftembed}{\mathbf{s}}
\newcommand{\queryembed}{\mathbf{q}}
\newcommand{\answerembed}{\mathbf{a}}

\newcommand{\uniondist}{\mathcal{U}}

\newcommand{\datathreshold}{\eta}

\begin{document}
	
\pagestyle{plain}
\pagenumbering{roman}
\maketitle



\begin{abstract}

\noindent Learning knowledge graph (KG) embeddings is an emerging technique for a variety of downstream tasks such as summarization, link prediction, information retrieval, and question answering. However, most existing KG embedding models neglect space and, therefore, do not perform well when applied to (geo)spatial data and tasks. For those models that consider space, most of them primarily rely on some notions of distance. These models suffer from higher computational complexity during training while still losing information beyond the relative distance between entities. In this work, we propose a location-aware KG embedding model called $\spexkge$. It directly encodes spatial information such as point coordinates or bounding boxes of geographic entities into the KG embedding space.  The resulting model is capable of handling different types of spatial reasoning. We also construct a geographic knowledge graph as well as a set of geographic query-answer pairs called $\dbgeo$  
to evaluate the performance of $\spexkge$ in comparison to multiple baselines. Evaluation results show that $\spexkge$ outperforms these baselines on the $\dbgeo$ dataset for geographic logic query answering task. This demonstrates the effectiveness of our spatially-explicit model and the importance of considering the scale of different geographic entities. Finally, we introduce a novel downstream task called \textit{spatial semantic lifting} which links an arbitrary location in the study area to entities in the KG via some relations. Evaluation on $\dbgeo$ shows that our model outperforms the baseline by a substantial margin.


\end{abstract} \pagebreak
\pagenumbering{arabic}

\section{Introduction and Motivation} \label{sec:intro}

\noindent The term \textit{Knowledge Graph} typically refers to a labeled and directed multi-graph of statements (called triples) about the world. These triples often originate from heterogeneous sources across domains. 
According to \citet{nickel2015review}, most of the widely used knowledge graphs are constructed in a curated (e.g., WordNet), collaborative (e.g., Wikidata, Freebase), or auto semi-structured (e.g., YAGO \citep{hoffart2013yago2}, DBpedia, Freebase) fashion rather than an automated unstructured approach (e.g., Knowledge Vault \citep{dong2014knowledge}). 
Despite containing billions of statements, these knowledge graphs suffer from \textit{incompleteness and sparsity} 
\citep{lao2011random, dong2014knowledge, mai2019contextual}. 
To address these problems, many relational machine learning models \citep{nickel2015review} have  been developed for knowledge graph completion tasks including several embedding-based techniques such as  
RESCAL \citep{nickel2012factorizing}, 
TransE \citep{bordes2013translating}, 
TransH \citep{wang2014knowledge},
HOLE \citep{nickel2016holographic},
R-GCN \citep{schlichtkrull2018modeling},
and TransGCN \citep{cai2019transgcn}. The key idea of the embedding-based technique \citep{bordes2013translating,wang2014knowledge,nickel2016holographic,cai2019transgcn, wang2017knowledge} is to project entities and relations in a knowledge graph onto a continuous vector space such that entities and relations can be quantitatively represented as vectors/embeddings.

The aforementioned incompleteness and sparsity problems also affect the performance of downstream tasks such as question answering \citep{wang2018towards} since missing triples or links result in certain questions
becoming unanswerable \citep{rajpurkar2018know}. 
Consequently, researchers have recently focused on relaxing these unanswerable queries or predicting the most probable answers based on knowledge graph embedding models \citep{wang2018towards,hamilton2018embedding,mai2019relaxing}.

Most research on knowledge graph embeddings has neglected spatial aspects such as the location of geographic entities despite the important role such entities play within knowledge graphs \citep{janowicz2012geospatial}. In fact, most of the current knowledge graph embedding models 
(e.g. TransE, TransH, TransGCN, R-GCN, and HOLE)
 ignore triples that contain \textit{datatype  properties}, and, hence, literals for dates, texts, numbers, geometries, and so forth. Put differently, properties such as \texttt{dbo:elevation}, \texttt{dbo:populationTotal}, and \texttt{dbo:areaWater} to name but a few are not considered during the training phase. 
Instead, these models strictly focus on triples with \textit{object type properties}, leading to substantial information loss in practice. 
A few models do consider a limited set of datatypes. 
LiteralE \citep{kristiadi2019incorporating} is one example, which encodes numeric and date information into its embedding space, while MKBE \citep{pezeshkpour2018embedding} encodes images and unstructured texts.
Therefore, in this work, we propose a novel technique which directly encodes spatial footprints, namely point coordinates and bounding boxes, thereby making them available while learning knowledge graph embeddings.

Geographic information forms the basis for many KG downstream tasks such as geographic knowledge graph completion \citep{qiu2019knowledge}, geographic ontology alignment \citep{zhu2016spatial}, geographic entity alignment \citep{trisedya2019entity}, geographic question answering \citep{mai2019relaxing}, and geographic knowledge graph summarization \citep{yan2019spatially}. 
In the following, we will focus on geographic logic query answering as an example and more concretely on conjunctive graph queries (CGQ) or logic queries \citep{hamilton2018embedding}. Due to the sparsity of information in knowledge graphs, many (geographic) queries are unanswerable 
\textit{without spatial or non-spatial reasoning}. 
Knowledge graph embedding techniques have, therefore, been developed to handle unanswerable questions \citep{hamilton2018embedding,wang2018towards,mai2019relaxing,mai2019contextual} by inferring new triples in the KG embedding space based on existing ones. 
However, since most KG embedding models cannot handle \textit{datatype  properties} thus cannot encode geographic information into the KG embedding space, they perform spatial reasoning tasks poorly in the KG embedding space, which in turn leads to a poor performance of handling unanswerable geographic questions.

\vspace*{1.0cm}
\hspace*{-\parindent}
\begin{minipage}[c]{\columnwidth}
	\begin{lstlisting}[
	linewidth=\columnwidth,
	breaklines=true,
	basicstyle=\ttfamily, 
	captionpos=b, 
	caption={Query $\queryA$: An unanswerable SPARQL query over DBpedia which includes a partonomy relation
	}, 
	label={q:part},
	frame=single
	]
SELECT ?State WHERE {
 ?RiverMouth dbo:state ?State.                    (a)
 ?River dbo:mouthPosition ?RiverMouth.               (b)
 ?River dbp:etymology dbr:Alexander_von_Humboldt.    (c)
}
	\end{lstlisting}
\end{minipage}
One example of unanswerable geographic questions that can be represented as a logic query is \textit{which states contain the mouth of a river which is named after Alexander von Humboldt?} (Query $\queryA$).
The corresponding SPARQL query is shown in Listing \ref{q:part}. Running this query against the DBpedia SPARQL endpoint yields no results. In fact, two rivers are named after von Humboldt - \texttt{dbr:Humboldt\_River} and \texttt{dbr:North\_Fork\_Humboldt\_River} - and both have mouth positions as entities in DBpedia (\texttt{dbr:Humboldt\_River\_\_mouthPosition\_\_1} and \texttt{dbr:North\_Fork\_Humboldt\_River\_\_sourcePosition\_\_1}). However, the \texttt{dbo:state} (or \texttt{dbo:isPartOf}) relation between these river mouths and other geographic features such as states is missing. This makes Query $\queryA$ unanswerable (graph query pattern (a) in Listing \ref{q:part}).
If we use the locations of the river mouths to perform a simple point-in-polygon test against the borders of all states in the US, we can deduce that \texttt{dbr:Nevada} contains both river mouths.

\vspace*{1.0cm}
\hspace*{-\parindent}
\begin{minipage}[c]{\columnwidth}
	\begin{lstlisting}[
	linewidth=\columnwidth,
	breaklines=true,
	basicstyle=\ttfamily, 
	captionpos=b, 
	caption={Query $\queryB$: A SPARQL query over DBpedia which indicates a simple point-wise distance relation 
	}, 
	label={q:dis},
	frame=single
	]
SELECT ?place WHERE {
 dbr:Yosemite_National_Park dbo:nearestCity ?place.
}
	\end{lstlisting}
\end{minipage}
Another example is the query in Listing \ref{q:dis}, which asks for \textit{the nearest city to Yosemite National Park} (Query $\queryB$). If the triple \texttt{dbr:Yosemite\_National\_Park} \texttt{dbo:nearestCity} \texttt{dbo:Mariposa,\_California} is missing from the current knowledge graph, Query $\queryB$ becomes unanswerable while it could simply be inferred by a  distance-based query commonly used in GIS. Similar cases can include cardinal directions such as \texttt{dbp:north}. 
 All these observations lead to the following research question: how could we enable spatial reasoning via  \textit{partonomic relations}, \textit{point-wise metric relations}, and \textit{directional relations} in the KG embedding-based systems?

One may argue that classical spatial reasoning can be used instead of direct location encoding to obtain answers to aforementioned questions. This is partially true for data and query endpoints that support GeoSPARQL and for datasets that are clean and complete. 
However, in some cases even GeoSPARQL-enabled query endpoints cannot accommodate spatial reasoning due to inherent challenges of representing spatial data in knowledge graphs. These challenges stem from principles of conceptual vagueness and uncertainty \citep{regalia2019computing}, and are further complicated by technical limitations.
In this study we aim at enabling the model to perform \textit{implicit spatial reasoning} in the hidden embedding space. Instead of performing classical spatial reasoning by explicitly carrying out spatial operations during query time, the spatial information (points or bounding boxes) of geographic entities (e.g., $Indianapolis$) are directly encoded into the entity embeddings which are jointly optimized with relation embeddings (e.g, $isPartOf$). The trained embeddings of geographic entities encode their spatial information while by embedding the spatial relations, we also hope to capture some of their implicit semantics for simple spatial reasoning tasks.
At query time, a normal link prediction process can be used to answer geographic questions and no explicit spatial reasoning is needed. 
Find more detail of this example in Section \ref{sec:exp}.

Existing approaches are only able to incorporate spatial information into the KG embedding space in a very limited fashion, e.g., through their training procedures.
Furthermore, they estimate entity similarities based on some form of distance measures among entities, and ignore their absolute positions or relative directions.
For example, \citet{trisedya2019entity} treated geographic coordinates as strings (a sequence of characters) and used a compositional function to encode these coordinate strings for geographic entities alignment.
In order to incorporate distance relations between geographic entities, both \citet{mai2019relaxing} and \citet{qiu2019knowledge} borrowed the translation assumption from TransE \citep{bordes2013translating}.
For each geographic triple $s = (h, \rel, t)$ in the KG, where $h$ and $t$ are geographic entities, the geospatial distance between $h$ and $t$ determines the frequency of resampling this triple such that triples containing two closer geographic entities are sampled more frequently, and thus these two geographic entities are closer in the embedding space.
Similarly, \citet{yan2019spatially} used distance information to construct virtual spatial relations between geographic entities during the knowledge graph summarization process. 
This data conversion process (coordinates to pairwise distances) is unnecessarily expensive and causes information loss, e.g., absolute positions and relative directional information. 
In this work, we explore to directly encode entity locations into a high dimensional vector space, which preserves richer spatial information than distance measures. These location embeddings can be trained jointly with knowledge graph embedding.

Location encoders \citep{mac2019presence,chu2019geo,mai2020multiscale} refer to the neural network models which encode a pair of coordinates into a high dimensional embedding which can be used in downstream tasks such as geo-aware fine-grained image classification \citep{mac2019presence,chu2019geo,mai2020multiscale} and Point of Interest (POI) type classification \citep{mai2020multiscale}. 
\citet{mai2020multiscale} showed that
multi-scale grid cell representation   outperforms commonly used   kernel based methods (e.g., RBF) as well as the single scale location encoding approaches.
Given the success of location encoding 
in other machine learning tasks, 
the question is whether we can incorporate the location encoder architecture into a knowledge graph embedding model to make it spatially explicit \citep{mai2019relaxing}. 
One initial idea is directly using a location encoder as the entity encoder which encodes the spatial footprint (e.g., coordinates) of a geographic entity into a high dimensional vector. Such entity embeddings can be 
used in different decoder architectures for different tasks. 
However, several challenges remain to be solved for this initial approach.

First, point location encoding  can handle  point-wise metric relations such as distance (e.g., \texttt{dbo:nearestCity}) as well as directional relations (e.g., \texttt{dbp:north}, \texttt{dbp:south}) in knowledge graphs, but it is not 
easy to encode regions which are critical for 
relations such as containment  (e.g., \texttt{dbo:isPartOf}, \texttt{dbo:location}, \texttt{dbo:city}, \texttt{dbo:state}, and \texttt{dbo:country}). 
For example, in Query $\queryA$, the location encoder can encode \texttt{dbr:Yosemite\_National\_Park} and \texttt{dbo:Mariposa,\_California} as two high dimensional embeddings based on which distance relations can be computed since the location embeddings preserve the relative distance information between locations \citep{mai2020multiscale}. 
However, point locations and location embeddings are insufficient to capture more complex relations between geographic entities such as containment as these require more complex spatial footprints (e.g., polygons). This indicates that we need to find a way to represent geographic entities as \textit{regions} instead of points in the embedding space based on location encoders, especially for large scale geographic entities such as \texttt{dbr:California}, which is represented as a single pair of coordinates (a point) in many widely used KGs. We call this \textit{scale effect} to emphasize the necessity of encoding the spatial extents of geographic entities instead of points, especially for large scale geographic entities.

The second challenge is how to seamlessly handle geographic and non-geographic entities together in the same entity encoder framework.
Since location encoder is an essential \textit{component} of the entity encoder, how should we deal with non-geographic entities that do not have spatial footprints? 
This is a non-trivial problem. For example, in order to weight triples using distance during KG embedding training, \citet{qiu2019knowledge} constructed a geographic knowledge graph which only contains geographic entities. \citet{mai2019relaxing} partially solved the problem by using a lower bound $l$ as the lowest triple weight to handle non-geographic triples. However, this mechanism cannot distinguish triples involving both geographic and non-geographic entities from triples that only contain non-geographic entities.

The third challenge is how to capture the spatial and other semantic aspects at the same time when designing spatially explicit KG embedding model based on location encoders. The embedding of a geographic entity is expected to capture both its spatial (e.g., spatial extent) and other semantic information (e.g., type information) since both of them are necessary to answer geographic questions. Take Query $\queryA$ in Listing \ref{q:part} as an example. Intuitively, to answer this query, the spatial information is necessary to perform partonomical reasoning to select geographic entities which \textit{contain} a given river mouth, while type information is required to filter the answers and get entities with type $state$. Therefore, we need both spatial and type information encoded in the entity embeddings to answer this question. The traditional KG embedding models fail to capture the spatial information which leads to a lower performance in geographic question answering.

Finally, thanks to the inductive learning nature of the location encoder, another interesting question is how to design a spatially-explicit KG embedding model so that it can be used to infer new relations between entities in a KG and any arbitrary location in the study area. We call this task \textbf{spatial semantic lifting} as an analogy to traditional \textit{semantic lifting} which refers to the process of associating unstructured content to semantic knowledge resources \citep{de2008semantic}. For example, given any location $\bx_{i}$, we may want to ask \textit{which radio station broadcasts at $\bx_{i}$}  i.e., to infer \texttt{dbo:broadcastArea}. None of the existing KG embedding models can solve this task.

 In this work, we develop a spatially-explicit knowledge graph embedding model, \spexkge,  which directly solves those challenges. \textbf{The contributions of our work are as follow}:

\begin{enumerate}
	\item We develop a spatially-explicit knowledge graph embedding model ($\spexkge$), which applies a location encoder to incorporate spatial information (coordinates and spatial extents) of geographic entities. To the best of our knowledge, this is the first KG embedding model that can incorporate spatial information, especially spatial extents, of geographic entities into the model architecture.
	\item $\spexkge$ is extended to an end-to-end geographic logic query answering model which predicts the most probable answers to unanswerable geographic logic queries over KG.
	\item We apply $\spexkge$ on a novel task called \textit{spatially semantic lifting}.
	Evaluations show that our model can substantially outperform the baseline by 9.86\% on AUC and 9.59\% on APR for the $\dbgeo$ dataset. Furthermore, our analysis shows that this model can achieve implicit spatial reasoning for different types of spatial relations.

\end{enumerate}

The rest of this paper is structured as follow. We briefly summarize related work in Section \ref{sec:relatedwork}. Then basic concepts are discussed in Section \ref{sec:concept}. In Section \ref{sec:problem}, we formalize the query answering and spatial semantic lifting task. Then, in Section \ref{sec:qa_background}, we give an overview of the logic query answering task before introducing our method. Section \ref{sec:method} describes the $\spexkge$ architecture. Experiments and evaluations are summarized in Section \ref{sec:exp}. Finally, we conclude our work in Section \ref{sec:conclusion}.

 \section{Related Work} \label{sec:relatedwork}
In this section, we briefly review related work on knowledge graph embeddings,  query answering, and location encoding.

\subsection{Knowledge Graph Embedding}  \label{subsec:kge_related}
Learning knowledge graph embeddings (KGE) is an emerging topic in both the Semantic Web and machine learning fields. The idea is to represent entities and relations as vectors or matrices within an embedding spaces such that these distributed representations can be easily used in downstream tasks such as KG completion and question answering. Many KG embedding models have been proposed such as RESCAL \citep{nickel2012factorizing}, 
TransE \citep{bordes2013translating}, and
TransH \citep{wang2014knowledge}. Most of these approaches cannot handle triples with data type properties nor triples involving spatial footprints.

The only KG embedding methods considering distance decay  between geographic entities are \citet{qiu2019knowledge} and \citet{mai2019relaxing}.
\citet{mai2019relaxing} computed the weight of each geographic triple $s = (h, \rel, t)$ as $\max(\ln{D \over dis(h,t) + \varepsilon}, l)$ where $h$ and $t$ are geographic entities, and $D$ is the longest (simplified) earth surface distance. $\varepsilon$ is a hyperparameter to avoid zero denominator and $l$ is the lowest edge weight we allow for each triple. As for non-geographic triples, $l$ is used as the triple weight. Then this knowledge graph is treated as an undirected, unlabeled, edge-weighted multigraph. An edge-weighted PageRank is applied on this multigraph. The PageRank score for each node/entity captures the the structure information of the original KG as well as the distance decay effect among geographic entities. These scores are used in turn as weights to sample the entity context from the 1-degree neighborhood of each entity which is used in the KG embedding training process. As for \citet{qiu2019knowledge}, the distance decay effect was deployed in a triple negative sampling process. Given a triple $s = (h, \rel, t)$ in the KG, each negative triple $s^{\prime} = (h^{\prime}, \rel, t^{\prime})$ of it was assigned a weight based on $w_{geo} = \dfrac{1}{1 + \Bigl\vert log_{10} \dfrac{dis(h,t)+\theta}{dis(h^{\prime},t^{\prime})+\theta} \Bigr\vert}$ where $\theta$ is a hyperparameter to avoid a zero denominator. $w_{geo}$ is used in the max-margin loss function for the embedding model training. Note that non-geographic triples are not considered in \citet{qiu2019knowledge}.
We can see that, instead of directly encoding an entity's location, they rely on some form of distance measures as weights for triple resampling. This process is computationally expensive and does not preserve other spatial properties such as direction. In contrast, our work introduces a direct encoding approach to handle spatial information.

\subsection{Query Answering}  \label{subsec:qa_related}
Compared to link prediction \citep{bordes2013translating}, query answering \citep{wang2018towards,hamilton2018embedding,mai2019contextual} focuses on a more complex problem since answering a  query requires a system to consider multiple triple patterns together. \citet{wang2018towards} designed an algorithm to answer a subset of SPARQL queries based on a pretrained KG embedding model. However, this is not an end-to-end model since the KG embedding training and query answering process are separated. \citet{hamilton2018embedding} proposed an end-to-end logic query answering model, $GQE$, which can answer conjunctive graph queries. $CGA$ \citep{mai2019contextual} further improved $GQE$ by using a self-attention based intersection operator. In our work, we will utilize $GQE$ and $CGA$ \citep{mai2019contextual} as the underlying logic query answering baseline. We provide an overview about logic query answering in Section \ref{sec:qa_background}.

\subsection{Location Encoding}  \label{subsec:locenc_related}
Generating representations of points/locations that can benefit representation learning is a longstanding problem in machine learning. There are many well-established methods such as the \textit{Kernel trick} \citep{scholkopf2001kernel} widely used in SVM classification and regression. However, these location representation methods use the positions of training examples as the centers of Gaussian kernels and thus need to memorize the training examples.

\citet{kejriwal2017neural} proposed a graph embedding approach to representing GeoNames locations as high dimensional embeddings. They converted the locations in GeoNames into a weighted graph where locations are nodes and the weight of each edge is computed based on the distance between two locations. Then a Glove \citep{pennington2014glove} word embedding model is applied on this generated graph to obtain the embedding for each location. Despite its novelty, this model is a transductive learning based model which means if new locations are added, the weighted graph has to be regenerated and the whole model needs to be retrained. In other words, this embedding approach can not be easily generalized to unseen locations. This calls for inductive learning \citep{hamilton2017inductive} based models.

Recently, location encoding technique \citep{mac2019presence,chu2019geo,mai2020multiscale} has been proposed to directly encode a location (a pair of coordinates) $\bx$ as a high-dimension vector which can be incorporated into multiple downstream tasks. As shown by \citet{mai2020multiscale}, the advantages of location encoding is that 
\textbf{1)} it can preserve  absolute position information as well as  relative distance and direction information between locations; 
\textbf{2)} it does not need to memorize the positions of training examples as all kernel based methods do \citep{scholkopf1997comparing}; 
\textbf{3)} In contrast to many transductive learning models, it is an inductive learning model \citep{battaglia2018relational} which can encode any location/point no matter it appears in the training dataset or not.

In theory, we can adopt any location encoder \citep{chu2019geo,mac2019presence,mai2020multiscale} to capture the spatial information of each geographic entity $\ent_{i}$ in a knowledge graph $\kg$. In this work, we utilize the $Space2Vec$ \citep{mai2020multiscale} location encoder, which is inspired by Nobel Prize-winning neuroscience research about grid cells \citep{abbott2014nobel} as well as the position encoding module of the Transformer model \citep{vaswani2017attention}. $Space2Vec$ first encodes a location $\bx$ as a multi-scale periodic representation $PE(\bx)$ by using sinusoidal functions with different frequencies and then feeds the resulting embedding into a $N$ layer feed forward neural network $\pemlp()$.
\begin{equation}
    \locenc(\bx) = \pemlp(PE(\bx))
    \label{equ:locenc}
\end{equation}

The advantages of such location encoder compared to previous work \citep{chu2019geo,mac2019presence} are that
\textbf{1)} it can be shown that location embeddings from  $Space2Vec$ are able to preserve global position information as well as relative distance and direction, and that
\textbf{2)}  multi-scale representation learning approach outperforms traditional kernel-based methods (e.g., RBF) as well as single-scale location encoding approaches \citep{chu2019geo,mac2019presence} for several machine learning tasks. In the following, we will use $\locenc()$ to denote the $Space2Vec$ model.

 \section{Basic Concepts} \label{sec:concept}
\begin{definition}[Geographic Knowledge Graph]
    \label{def:kg} 
    A geographic knowledge graph $\kg = (\entset, \edgeset)$ is a directed edge and node labeled multigraph where $\entset$ is a set of entities/nodes and $\edgeset$ is the set of directed edges. Any directed and labeled edge will be called a triple $\triple = (h,\rel,t)$ where the nodes become heads $h \in  \entset$ and tails $t \in  \entset$, and the role label $\rel \in \relset$ will be called the relationship between them. The set of triples/statements contained by $\kg$ is denoted as $\triset$ and $\relset$ denotes as the set of relations (predicates, edge labels) in $\kg$. Each triple can also be represented as $\rel(h,t)$, or $\rel^{-1}(t, h)$ where $\rel^{-1}$ indicates the inverse relation of $\rel$. $\domain(\rel)$ and $\range(\rel)$ indicate the domain and range of relation $\rel$.

    $\typefun(): \entset \to \typeset $ is a function which maps an entity $\ent \in \entset$ to a unique type $\type  \in \typeset$, where $\typeset$ is the set of all entity types in $\kg$.\footnote{
    Note that, in many knowledge graphs (e.g., DBpedia, Wikidata), an entity can belong to multiple types. We use this definition to be in line with many existing work \citep{hamilton2018embedding,mai2019contextual} so that we can compare our results. It is easy to relax this requirement which we will discuss in Section \ref{subsec:entemb}.
    }
    
    The geographic entity set $\ptentset$ is a subset of $\entset$ ($\ptentset \subseteq \entset$). $\ptfun(\cdot)$ is a mapping function that maps any geographic entity $\ent \in \ptentset$ to its geographic location (coordinates) $\ptfun(\ent) = \bx$ where 
    $\bx \in \studyarea  \subseteq \Real^{2}$. Here $\studyarea $ denotes the bounding box containing all geographic entities in the studied knowledge graph $\kg$. We call it study area.
    
    $\pgonentset$ is a subset of $\ptentset$ ($\pgonentset \subseteq \ptentset$) which represents the set of large-scale geographic entities whose spatial extent cannot be ignored. In this work, we use a bounding box to represent
    a geographic entity's spatial footprint. $\pgonfun(\cdot)$ is a mapping function defined on $\pgonentset$ that maps a geographic entity $\ent \in \pgonentset$ to its spatial extent  $\pgonfun(\ent)$ and $ \pgonfun(\ent) = [\bx^{min};\bx^{max}] \in \Real^{4}$. In the vector concatenation above, $\bx^{min}, \; \bx^{max} \in \studyarea  \subseteq \Real^{2}$ indicate the southwest and northeast point of the entity's bounding box.

\end{definition}

Note that in many existing knowledge graphs, a triple can include a datatype property (e.g., \texttt{dbo:abstract}) implying that the tail is a literal. In line with related work \citep{bordes2013translating,nickel2016holographic,nickel2015review,wang2017knowledge}, we do not consider this kind of triples here in general. However, we do consider datatype properties about the spatial footprints of geographic entities implicitly by using $\ptfun(\cdot)~or~ \pgonfun(\cdot)$.\footnote{It is worth mentioning that most KG to date merely store point geometries even for features such as the United States.} While we do not model them directly as triples, we use the spatial footprints of geographic entities as input features for the  entity encoder.

\begin{definition}[Conjunctive Graph Query (CGQ)]
	\label{def:cgq}
	A query $\cgq \in Q(\kg)$ that can be written as follows:
	\begin{eqnarray*}
		& \cgq = \variable_{?}. \exists \variable_{1}, \variable_{2},..,\variable_{m}: \gp_{1} \land \gp_{2} \land ... \land \gp_{n} \\
		where &\; \gp_{i} = \rel_{i}(\ent_{k}, \variable_{l}), \variable_{l} \in \{\variable_{?},  \variable_{1}, \variable_{2},..,\variable_{m}\}, \ent_{k} \in \entset, \rel \in \relset \\
		or& \; \gp_{i} = \rel_{i}(\variable_{k}, \variable_{l}), \variable_{k}, \variable_{l} \in \{\variable_{?},  \variable_{1}, \variable_{2},..,\variable_{m}\}, k \neq l, \rel \in \relset
		\label{equ:cgq}
	\end{eqnarray*}
\end{definition}

\begin{figure}[]
	\centering
	\setlength{\unitlength}{0.1\columnwidth}
	\begin{picture}(10,10)
    	\put(2.0,8){\includegraphics[width=0.63\columnwidth]{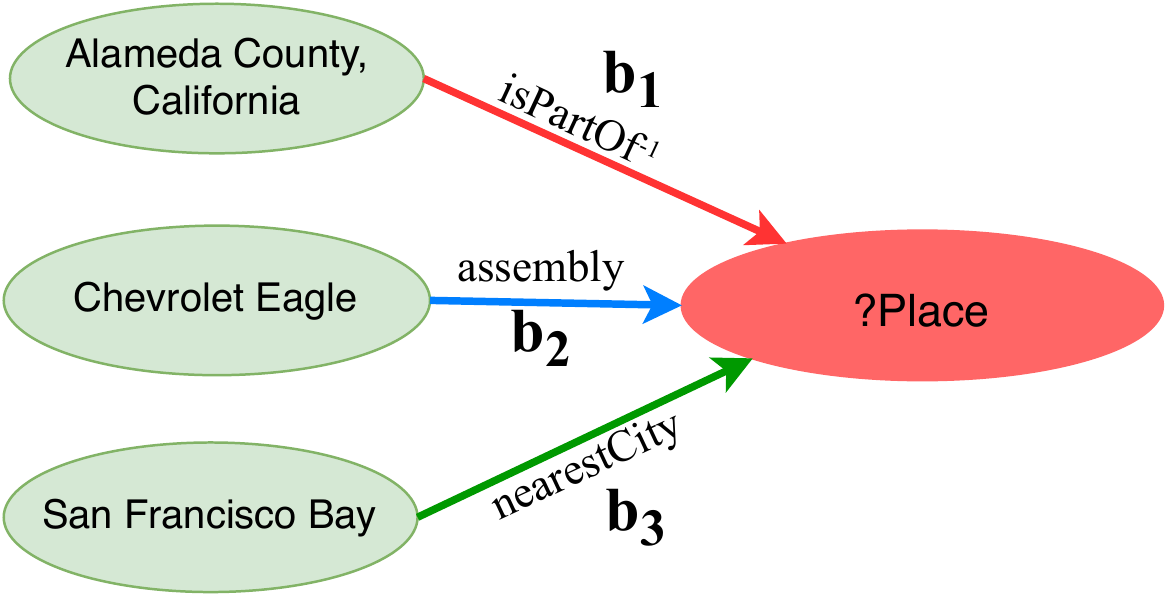}}

    	\put(1.5,7.0){$?Place : IsPartOf^{-1}(Alameda \; County, ?Place) \land $}
    	\put(2.5,6.6){$Assembly(Chevrolet \; Eagle, ?Place) \land $}
    	\put(2.5,6.2){$NearestCity(San \; Francisco \; Bay, ?Place) $}
    	\put(0,6.0){\framebox(10,5.9){}}
    	 \put(0,0){\includegraphics[width=1.0\columnwidth]{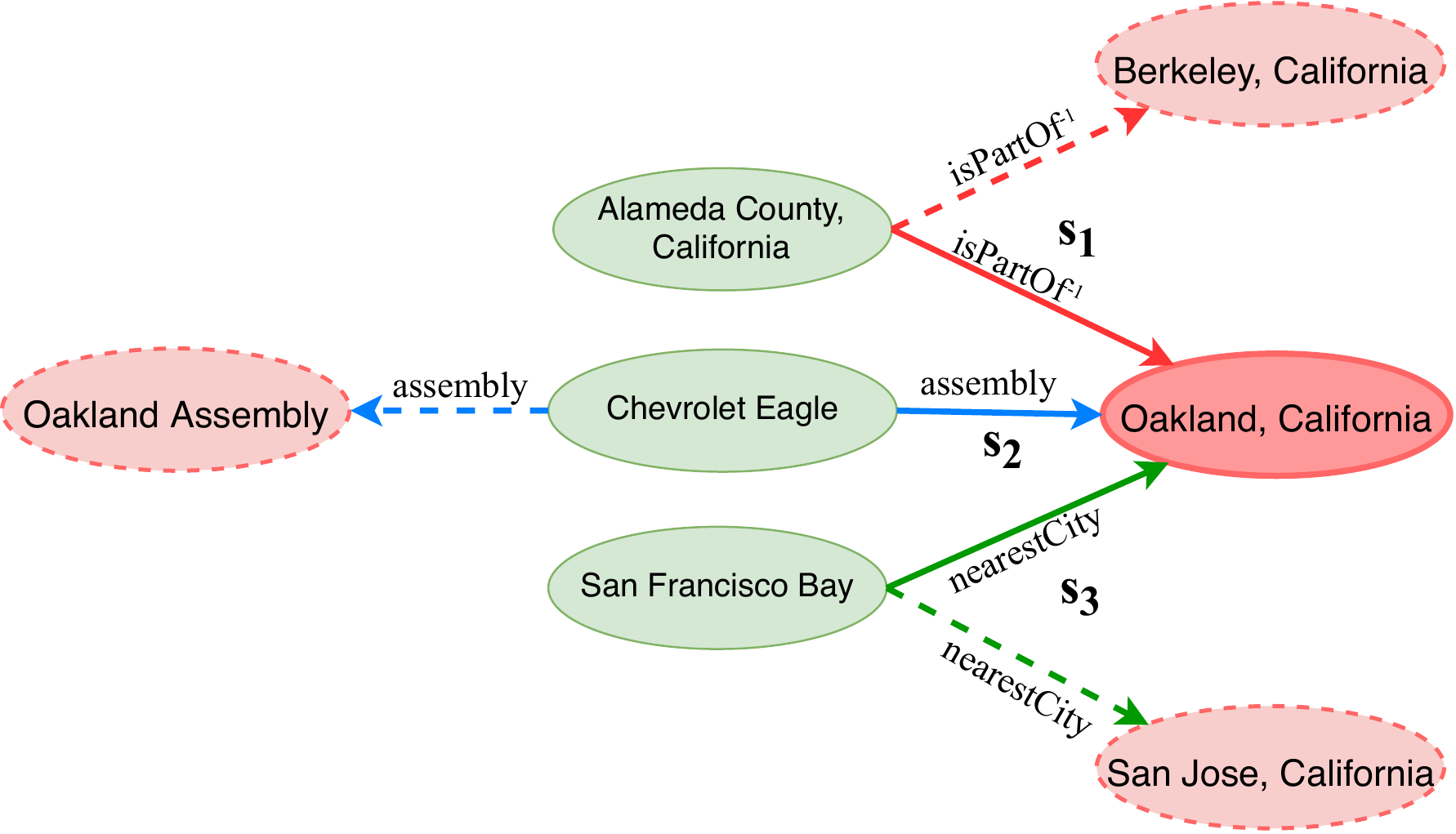}}
	\end{picture}
	\caption{
	\textbf{Query $\queryC$:}
	\textbf{Top box:}  
	Conjunctive Graph Query and Directed Acyclic Graph of the query structure corresponding to the SPARQL query in Listing \ref{q:gcgq}. $\bgpone$, $\bgptwo$, and $\bgpthree$ indicates three basic graph patterns in query $\queryC$.
 \texttt{?Place} is the target variable indicated as the red node while three green nodes 
 are anchor nodes. There is no bound variable in this query.
	\textbf{Below:}
	The matched underlining KG patterns represented by solid arrows.
	$\triple_{1}$, $\triple_{2}$, and $\triple_{3}$ indicates the matched triples for $\bgpone$, $\bgptwo$, and $\bgpthree$ respectively for query $\queryC$.
	}
	\label{fig:qg}
	\vspace{-0.45cm}
\end{figure}

Conjunctive graph queries are also called logic queries. Here $Q(\kg)$ is a set of all conjunctive graph queries that can be asked over $\kg$. $\variable_{?}$ denotes the target variable of query $\cgq$ (\textit{target node}) which will be replaced with the answer entity $\answer$, while $\variable_{1}, \variable_{2},..,\variable_{m}$ are existentially quantified bound variables (\textit{bound nodes}). $\{ \ent_{k} | \ent_{k} \; in \; \cgq \}$ is a set of anchor nodes and $\gp_{i}$ is a basic graph pattern in this CGQ. 
We define the dependency graph of $\cgq$ as the graph with basic graph pattern $\{\gp_{1}, ..., \gp_{n}\}$ formed between the anchor nodes $\{ \ent_{k} | \ent_{k} \; in \; \cgq \}$ and the variable nodes $\variable_{?}, \variable_{1}, \variable_{2},..,\variable_{m}$ (Figure \ref{fig:qg}). 
Each conjunctive graph query can be written as a SPARQL query.\footnote{For the detail comparison between CQG and SPARQL 1.1 query, please refer to Section 2.1 of \citet{mai2019contextual}.}

Note that the dependency graph of $\cgq$ represents computations on the  KG and is commonly assumed to be a \textit{directed acyclic graph (DAG)} \citep{hamilton2018embedding} where the entities (\textit{anchor nodes}) $\ent_{k}$ in $\cgq$ are the source nodes and the target variable $\variable_{?}$ is the unique sink node. This restriction makes the logic query answering task in line with the usual question answering set up (e.g., semantic parsing  \citep{berant2013semantic,liang2017neural}).

\begin{definition}[Geographic Conjunctive Graph Query (GCGQ)]
	\label{def:gcgq}
	A conjunctive graph query $\cgq \in Q(\kg)$ is said to be a geographic conjunctive graph query if the answer entity $\answer$ corresponding to the target variable $\variable_{?}$ is a geographic entity, i.e., $\answer = \qagoldfun(\kg, \cgq) \wedge \answer \in \ptentset$ where $\qagoldfun(\kg, \cgq)$ indicates the answer when executing query $\cgq$ on $\kg$. We denote all possible geographic CGQ on $\kg$ as $Q_{geo}(\kg) \subseteq Q(\kg)$.
\end{definition}

An example geographic conjunctive query $\queryC$ is shown in Figure \ref{fig:qg} whose corresponding SPARQL query is shown in Listing \ref{q:gcgq}. The  corresponding natural language question is \textit{[which city in Alameda County, California is the assembly place  of Chevrolet Eagle and the nearest city to San Francisco Bay]}. This query is especially interesting since it includes a non-spatial relation (\texttt{dbo:assembly}), a point-wise metric spatial relation (\texttt{dbo:nearestCity}) and a partonomy relation (\texttt{dbo:isPartOf}). 
Note that executing each basic graph pattern in Query $\queryC$ over DBpedia will yield multiple answers. For example, $\bgpone$ will return all subdivisions of Alameda County, California. $\bgptwo$ matches multiple assembly places of Chevrolet Eagle such as \texttt{dbr:Oakland,\_California}, \texttt{dbr:Oakland\_Assembly}, and \texttt{dbr:Flint,\_Michigan}. Interestingly, \texttt{dbr:Oakland\_Assembly} should be located in \texttt{dbr:Oakland,\_California} while there are no relationship between them in DBpedia except for their spatial footprints which can be inferred that they are closed to each other. $\bgpthree$ will return three entities\footnote{\texttt{dbo:nearestCity} triples in DBpedia are triplified from the ``Nearest major city'' row of the info box in each entity's corresponding Wikipedia page which may contain several cities. See \url{http://dbpedia.org/resource/San_Francisco_Bay}.} - \texttt{dbr:San\_Francisco}, \texttt{dbr:San\_Jose,\_California} and \texttt{dbr:Oakland,\_California}. Combining these three basic graph patterns will yield one answer \texttt{dbr:Oakland,\_California}. In our knowledge graph, both triple $\triple_{1}$ (See Figure \ref{fig:qg}) and triple $\triple_{2}$ are missing which makes Query $\queryC$ an unanswerable geographic query.

\vspace*{1.0cm}
\hspace*{-\parindent}
\begin{minipage}[c]{\columnwidth}
	\begin{lstlisting}[
	linewidth=\columnwidth,
	breaklines=true,
	basicstyle=\ttfamily, 
	captionpos=b, 
	caption={Query $\queryC$: A geographic conjunctive query which is rewritten as a SPARQL query over DBpedia including both non-spatial relations and different types of spatial relation.}, 
	label={q:gcgq},
	frame=single
	]
SELECT ?place  WHERE{
 ?place dbo:isPartOf dbr:Alameda_County,_California.     (1)
 dbr:Chevrolet_Eagle dbo:assembly ?place.                (2)  
 dbr:San_Francisco_Bay dbo:nearestCity ?place.           (3)
}
	\end{lstlisting}
\end{minipage}

 \section{Problem Statement}  \label{sec:problem}

In this work, we focus on 
two geospatial tasks - \textit{geographic logic query answering} and \textit{spatial semantic lifting}. 

\begin{problem}[Logic Query Answering]
	\label{def:gqa}
    Given a geographic knowledge graph $\kg$ 
    and an unanswerable conjunctive graph query $\cgq \in Q(\kg)$ (i.e., $\qagoldfun(\kg, \cgq) = \emptyset$), a query embedding function $\kgqefun_{\kg,\theta}(\cgq) : Q(\kg) \to \Real^\embdim$, which is parameterized by $\theta$, is defined to map $\cgq$ to a vector representation of $\embdim$ dimension. The most probable answer $\answer^{\prime}$ to $\cgq$ is the entity nearest to $\queryembed = \kgqefun_{\kg,\theta}(\cgq)$ in the embedding space:
    \begin{equation}
        \answer^{\prime} = \argmax_{\ent_{i} \in \entset} \;  \cosine(\kgqefun_{\kg,\theta}(\cgq), \enc(\ent_{i})) = \argmax_{\ent_{i} \in \entset} \; \cosine(\queryembed, \entemb_{i})
    \label{equ:qa}
    \end{equation}
\end{problem}

Here $\entemb_{i} = \enc(\ent_{i}) \in \Real^\embdim$ is the entity embedding of $\ent_{i}$ produced by an embedding encoder $\enc()$. $\cosine(\cdot)$ denotes the cosine similarity function:
\begin{equation}
\cosine(\queryembed, \entemb_{i}) = \dfrac{\queryembed \cdot \entemb_{i}}{\parallel \queryembed \parallel  \parallel \entemb_{i} \parallel}
\label{equ:cos}
\end{equation}

Note that $\cgq$ can be a geographic query or non-geographic query, i.e., $\cgq \in (Q(\kg) \setminus Q_{geo}(\kg))\vee Q_{geo}(\kg)$. \textit{Geographic logic query answering} indicates a logic query answering process over $Q_{geo}(\kg)$. The query embedding function $\kgqefun_{\kg,\theta}(\cgq)$ is constructed based on all three components of \spexkge~without any extra parameters: 
$\enc()$, 
$\proj()$, and 
$\inter()$, i.e., $\theta = \{\theta_{\enc}, \theta_{\proj}, \theta_{\inter}\}$.

\begin{problem}[Spatial Semantic Lifting]
	\label{def:spasemlift}
    Given a geographic knowledge graph $\kg$ and 
    an arbitrary location $\bx \in \studyarea \subseteq \Real^{2}$ from the current study area $\studyarea$, and a relation $\rel \in \relset$ such that $\domain(\rel) \subseteq \ptentset$, we define a spatial semantic lifting function $\spasemliftfun_{\kg,\theta_{ssl}}(\bx, \rel) : \studyarea \times \relset \to \Real^\embdim$, which is parameterized by $\theta_{ssl}$, to map $\bx$ and $\rel$ to a vector representation of $\embdim$ dimension, i.e., $\spasemliftembed = \spasemliftfun_{\kg,\theta_{ssl}}(\bx, \rel) \in \Real^\embdim$. A nearest neighbor search is utilized to search for the most probable entity $\spasemliftent \in \ptentset$ so that a virtual triple can be constructed between location $\bx$ and $\spasemliftent$, i.e., $\rel(\bx, \spasemliftent)$, where

    \begin{equation}
        \spasemliftent = \argmax_{\ent_{i} \in \entset} \;  \cosine(\spasemliftfun_{\kg,\theta_{ssl}}(\bx, \rel), \enc(\ent_{i})) = \argmax_{\ent_{i} \in \entset} \; \cosine(\spasemliftembed, \entemb_{i})
    \label{equ:spasemlift}
    \end{equation}
\end{problem}

The spatial semantic lifting function $\spasemliftfun_{\kg,\theta_{ssl}}(\bx, \rel)$ consists of two components of \spexkge~ without any extra parameter: $\enc()$ and $\proj()$, i.e., $\theta_{ssl} = \{\theta_{\enc}, \theta_{\proj}\}$. This spatial semantic lifting task is related to the link prediction task \citep{lao2011random} which is commonly used in the knowledge graph embedding literature \citep{bordes2013translating,nickel2016holographic,cai2019transgcn}. The main difference is that instead of predicting links between entities in the original knowledge graph $\kg$ as link prediction does, spatial semantic lifting links an arbitrary location $\bx$ to $\kg$. \textit{Since none of the existing KG embedding models can directly encode locations, they cannot be used for spatial semantic lifting}. \section{Logic Query Answering Backgrounds} \label{sec:qa_background}

Before introducing our $\spexkge$ model, we will first give an overview of how previous work \citep{hamilton2018embedding,mai2019contextual} tackled the logic query answering task with KG embedding models. 
Generally speaking, a logic query answering model is composed of three major components: entity encoder $\enc()$, projection operator $\proj()$, and intersection operator $\inter()$.
\begin{enumerate}
	\item \textbf{Entity encoder $\enc()$}: represents each entity as a high dimension vector (embedding);
	\item \textbf{Projection operator $\proj()$}: given a basic graph pattern $\gp = \rel(\ent_{i}, \variable_{j})$ (or $\gp = \rel(\variable_{i}, \variable_{j})$) in a CGQ $\cgq$, while the subject embedding $\entemb_{i}$ (or $\varemb_{i}$) of entity $\ent_{i}$ (or Variable $\variable_{i}$) is known beforehand, $\proj()$ \textit{projects} the subject embedding through a relation specific matrix to predict the embedding of $\variable_{j}$. 
	\item \textbf{Intersection operator $\inter()$}: integrates different predicted embeddings of the same Variable (e.g., $\variable_{j}$) from different basic graph patterns into one single embedding to represent this variable.
\end{enumerate}
Given these three neural network modules, any CGQ $\cgq$ can be encoded according by following their DAG query structures such that the embedding of the unique target variable $\variable_{?}$ for each query can be obtained - $\varemb_{?}$. We call it \textit{query embedding $\queryembed = \kgqefun_{\kg,\theta}(\cgq) = \varemb_{?}$} for CGQ $\cgq$. Then the most probable answer is obtained by a nearest neighbor search for $\queryembed$ in the entity embedding space (See Equation \ref{equ:qa}). Our work will follow the same model component setup and query embedding computing process. However, neither \citet{hamilton2018embedding} nor \citet{mai2019contextual} has considered encoding spatial information of geographic entities into the entity embedding space which is the core contribution of our work. Moreover, we extend the current model architecture such that it can also be applied to the spatial semantic lifting task. This new task cannot be handled by previous work. In the following, we will use $\cdot^{(GQE)}$ and $\cdot^{(CQA)}$ to indicate that these are model components used by \citet{hamilton2018embedding} and \citet{mai2019contextual}:

\subsection{Entity Encoder}
\label{subsec:embenc_bk}

In general, an entity encoder aims at representing any entity in a KG as a high dimension embedding so that it can be fed into following neural network modules. The normal practice shared by most KG embedding models \citep{bordes2013translating,nickel2016holographic,wang2014knowledge,schlichtkrull2018modeling,mai2018support,mai2019relaxing,cai2019transgcn,qiu2019knowledge} is to initialize an embedding matrix 
randomly where each column indicates an embedding for a specific entity. The entity encoding becomes an \textit{embedding lookup} process and these embeddings will be updated during the neural network backpropagation during training time.

Previous work has demonstrated that most of the information captured by entity embeddings is type information  \citep{hamilton2017representation,hamilton2018embedding}. So \citet{hamilton2018embedding} and \citet{mai2019contextual} took a step further and used a type-specific embedding lookup approach. We call the resulting module entity feature encoder $\featenc()$.

\begin{definition}[Entity Feature Encoder: $\featenc()$]
	\label{def:feenc}
	Given any entity $\ent_{i} \in \entset$  with type $\type_{i} = \typefun(e_{i}) \in \typeset$ from $\kg$, entity feature encoder $\featenc()$ computes the feature embedding $\feemb_{i} \in \Real^{\fedim}$ which captures the type information of entity $\ent_{i}$ by using an embedding lookup approach:
	\begin{equation}
	\feemb_{i} = \featenc(e_{i}) =  \dfrac{\embmat_{\type_{i}}\onehotvec_{i}^{(\type)}}{\parallel\embmat_{\type_{i}}\onehotvec_{i}^{(\type)}\parallel_{L2}}
	\label{equ:feemb}
	\end{equation}
\end{definition}

Here $\embmat_{\type_{i}} \in \realnum^{\fedim \times |\typeset|}$ is the type-specific embedding matrix for all entities with type $\type_{i} = \typefun(e_{i}) \in \typeset$. $\onehotvec_{i}^{(\type)}$ is a one-hot vector such that $\embmat_{\type_{i}}\onehotvec_{i}^{(\type)}$ will perform an embedding lookup operation which selects an entity feature embedding from the corresponding column. $\parallel\cdot\parallel_{L2}$ indicates the $L2$-norm.

\begin{figure}[h!]
	\centering
	\includegraphics[width=1.0\columnwidth]{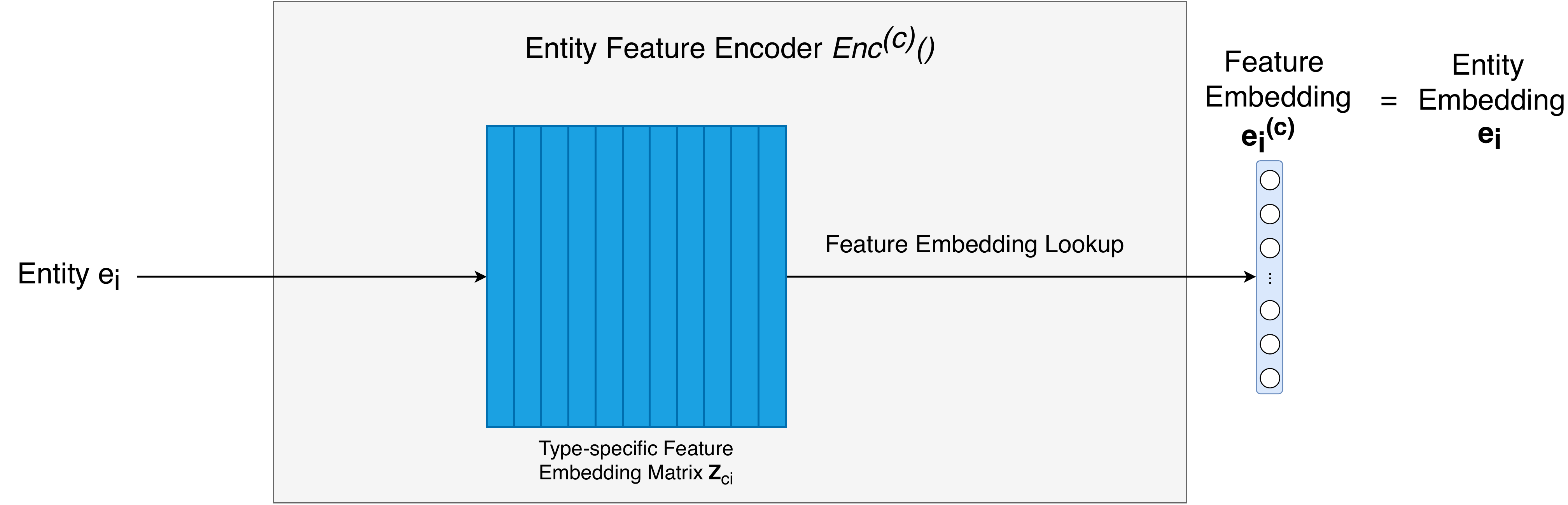}
	\caption{
	The entity encoder used by \citet{hamilton2018embedding} and \citet{mai2019contextual}.} 
	\label{fig:entenc_gqe}
\end{figure}

Both \citet{hamilton2018embedding} and \citet{mai2019contextual} use $\featenc()$ as their entity encoder (See Equation \ref{equ:ent_enc_gqe}). Figure \ref{fig:entenc_gqe} is an illustration of their approach. Note that this encoder does not consider the spatial information (e.g., coordinates and spatial extents) of geographic entities which causes a lower performance for answering geographic logic queries. As for our $\spexkge$ model, we add an additional entity space encoder $\spaenc()$ to handle this (See Definition \ref{def:peenc} ).

\begin{equation}
\encgqe(e_{i}) = \enccga(e_{i}) = \featenc(e_{i})
\label{equ:ent_enc_gqe}
\end{equation}

\subsection{Projection Operator}
\label{subsec:proj_bk}
The projection operator is utilized to do link prediction: given a basic graph pattern $\gp = \rel(\subject_{i}, \variable_{j})$ in a conjunctive graph query $\cgq$ with relation $\rel$ in which $\subject_{i}$ is either an entity $\ent_{i}$ (an anchor node in $\cgq$) or an existentially quantified bound variable $\variable_{i}$, the projection operator $\proj()$ predicts the embedding $\entemb_{i}^{\prime} \in \Real^{\fedim}$ for Variable $\variable_{j}$. Here, the embedding of $\subject_{i}$ can be either the entity embedding $\entemb_{i} = \featenc(\ent_{i})$ or the computed embedding $\varemb_{i}$ for $\variable_{i}$ which is known beforehand. Both \citet{hamilton2018embedding} and \citet{mai2019contextual} share the same projection operator $\projgqe = \projcga$ (See Equation \ref{equ:proj_gqe}) by using a bilinear matrix $\projmat_{\rel} \in \Real^{\fedim \times \fedim}$. $\projmat_{\rel}$ can also be a bilinear diagonal matrix as DisMult \citep{yang2014embedding} whose corresponding projection operator is indicated as $\projgqediag$.

\begin{equation}
\entemb_{i}^{\prime} = \begin{cases}
\projgqe(\ent_{i}, \rel) = \projcga(\ent_{i}, \rel) = \projmat_{\rel}\featenc(\ent_{i}) = \projmat_{\rel}\entemb_{i} \; & if \; input \; =  (\ent_{i}, \rel)    \\
\projgqe(\variable_{i}, \rel) = \projcga(\variable_{i}, \rel) = \projmat_{\rel}\varemb_{i} \; & if \; input \; =  (\variable_{i}, \rel)    
\end{cases}
\label{equ:proj_gqe}
\end{equation}

In $\spexkge$, we extend projection operator $\proj()$ so that it can be used in the spatial semantic lifting task (See Definition \ref{def:proj}).

\begin{figure}[h!]
	\centering
	\includegraphics[width=1.0\columnwidth]{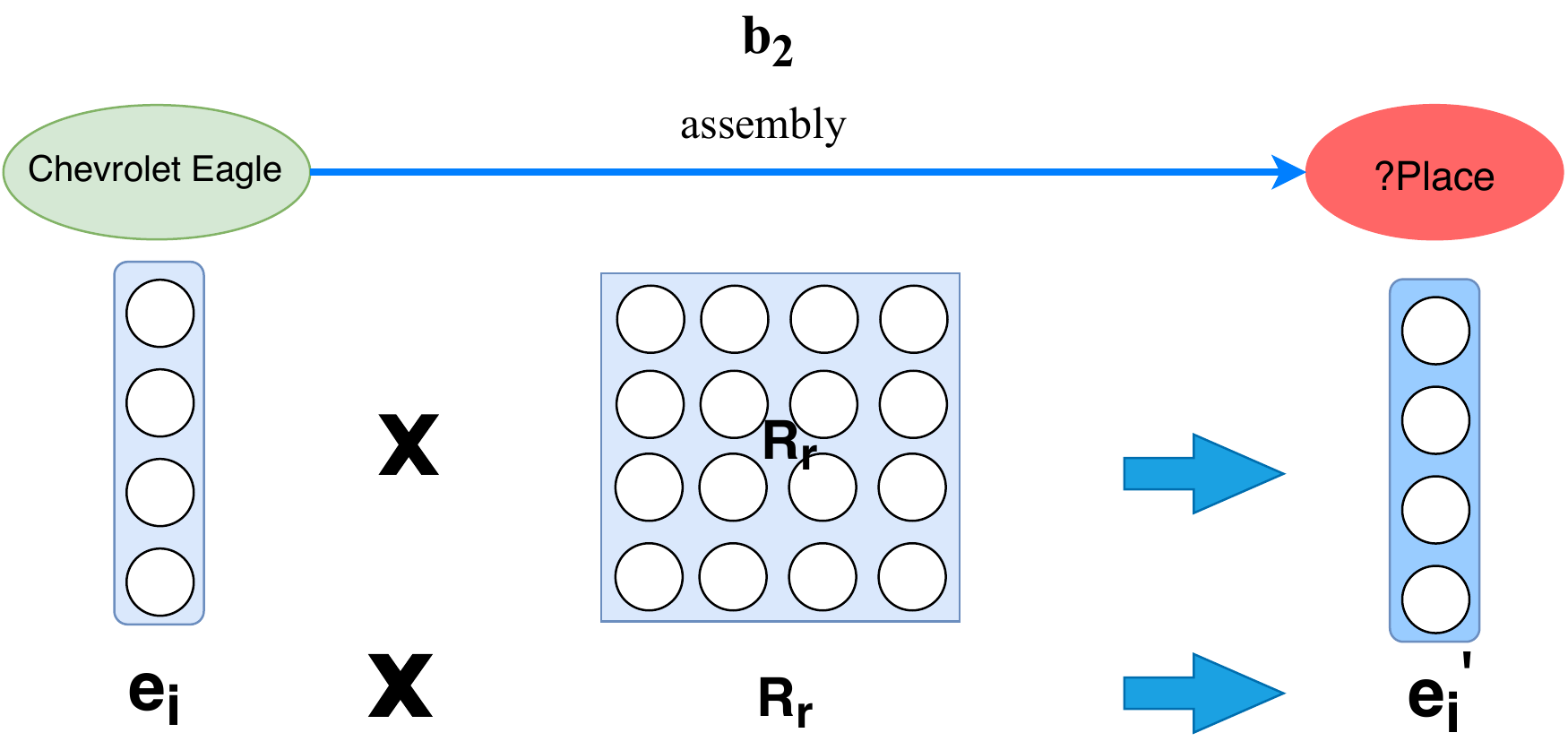}
	\caption{An illustration of projection operator $\projgqe() = \projcga()$ used by \citet{hamilton2018embedding} and \citet{mai2019contextual}.} 
	\label{fig:proj_gqe}
\end{figure}

Figure \ref{fig:proj_gqe} uses the basic graph pattern $\bgptwo = Assembly(Chevrolet \; Eagle, ?Place)$ in Figure \ref{fig:qg} as an example to demonstrate how to do link prediction with $\projgqe() = \projcga()$. The result embedding $\entemb_{?2}$ can be treated as the prediction of the embedding of Variable \texttt{?Place}. By following the same process, we can predict the embedding of the variable \texttt{?Place} from the other two basic graph patterns $\bgpone$ and  $\bgpthree$ - $\entemb_{?1}$ and $\entemb_{?3}$.

\subsection{Intersection Operator}  \label{subsec:inter}
\begin{figure}[h!]
	\centering
	\includegraphics[width=1.0\columnwidth]{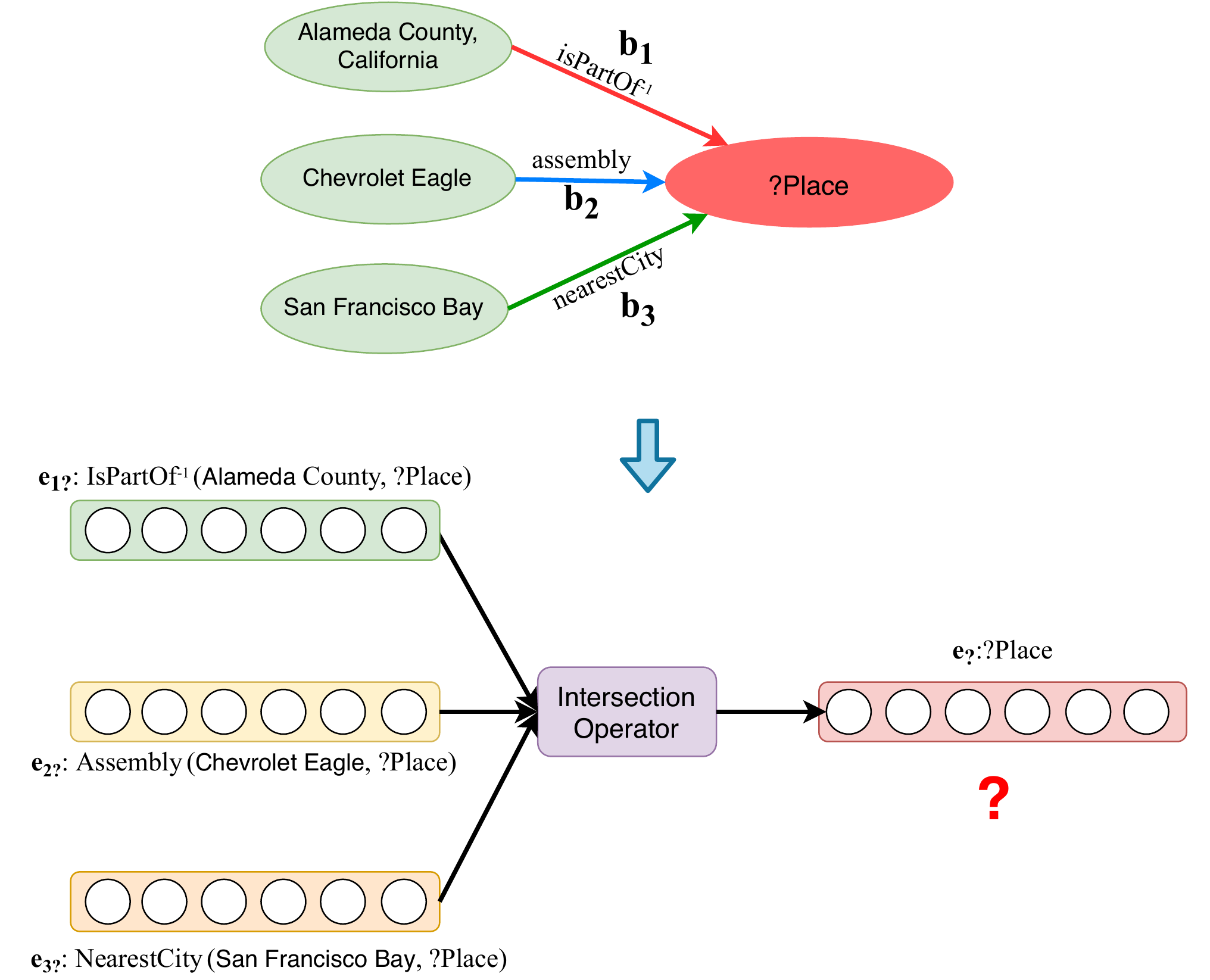}
	\caption{An illustration of intersection operator $\inter()$.} 
	\label{fig:inter}
\end{figure}

The intersection operator $\inter()$ is used to \textit{integrate} multiple embeddings $\entemb_{1?}$, $\entemb_{2?}$, ..., $\entemb_{i?}$,..., $\entemb_{\neisize ?}$ which represent the same (bound or target) variable $\variable_{?}$ in a CGQ $\cgq$ to produce one single embedding $\ent_{?}$ to represent this variable. Figure \ref{fig:inter} illustrates this idea by using CGQ $\queryC$ in Figure \ref{fig:qg} as an example where $\entemb_{?1}$, $\entemb_{?2}$ and $\entemb_{?3}$ indicates the predicted embedding of \texttt{?Place} from three different basic graph pattern $\bgpone$, $\bgptwo$, and $\bgpthree$. The intersection operator \textit{integrates} them into one single embedding $\entemb_{?}$ to represent \texttt{?Place}. Since \texttt{?Place} is the target variable of $\cgq$, $\entemb_{?}$ is the final \textit{query embedding} we use to do nearest neighbor search to obtain the most probable answer (See Task \ref{def:gqa}). More formally,

\begin{definition}[Intersection Operator $\inter()$]
    \label{def:inter}
    Given a set of $\neisize$ different input embeddings $\entemb_{1?}$, $\entemb_{2?}$, ..., $\entemb_{j?}$,..., $\entemb_{\neisize ?}$ 
    , intersection operator $\inter()$ produces one single embedding $\entemb_{?}$: 
    \begin{equation}
		\entemb_{?} = \inter(\{ \entemb_{1?}, \entemb_{2?}, ..., \entemb_{j?},..., \entemb_{\neisize ?} \})
		\label{equ:inter}
	\end{equation}
\end{definition}

Intersection operator $\inter()$ represents the logical conjunction in the embedding space. Any permutation invariant function can be used here as a conjunction such as elementwise mean, maximum, and minimum.  We can also use any permutation invariant neural network architecture \citep{zaheer2017deep} such as Deep Sets \citep{zaheer2017deep}.  $\gqe$ \citep{hamilton2018embedding} used an elementwise minimum plus a feed forward network as the intersection operator which we indicate as $\intergqe()$. \citet{mai2019contextual} showed that their $\cga$ model with a self-attention based intersection operator $\intercga()$ can outperform $\gqe$. So in this work, we use $\intercga()$ as the intersection operator $\inter()$. Readers that are interested in this technique are suggested to check \citet{mai2019contextual} for more details.

\subsection{Query Embedding Computing}  \label{subsubsec:queryemb}
\citet{hamilton2018embedding} proposed a way to compute the query embedding of a CGQ $\cgq$ based on these three components. Given a CGQ $\cgq$, we can encode all its anchor nodes (entities) into entity embedding space using $\enc()$. Then we recursively apply the projection operator $\proj()$ and intersection operator $\inter()$ by following the DAG of $\cgq$ until we get an embedding for the target node (variable $\variable_{?}$), i.e., $\queryembed = \kgqefun_{\kg,\theta}(\cgq) = \varemb_{?}$. Then we use the nearest neighbor search in the entity embedding space to find the \textit{closest} embedding, whose corresponding entity will be the predicted answer to Query $\cgq$. For details of the query embedding algorithm, please refer to \citet{hamilton2018embedding}.

\begin{figure}[h!]
	\centering
	\includegraphics[width=1.0\columnwidth]{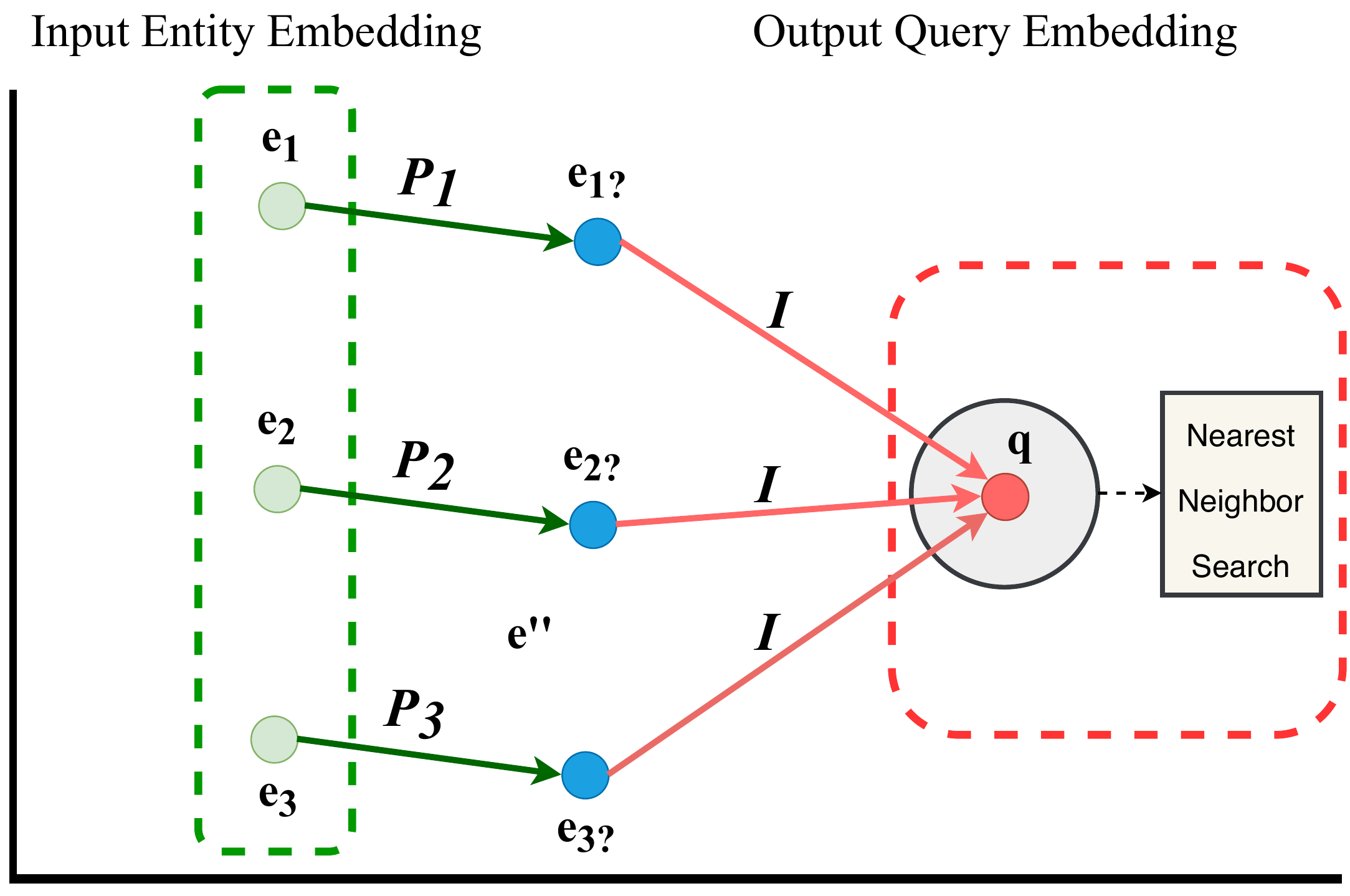}
	\caption{An illustration of (geographic) logic query answering in the embedding space} 
	\label{fig:qa}
\end{figure}

Figure \ref{fig:qa} gives an illustration of the query embedding computation process in the embedding space by using Query $\queryC$ as an example. We first use $\enc()$ to get the embeddings of three anchor nodes (see the dash green box in Figure \ref{fig:qa}.). Then $\proj()$ (three green arrows) is applied to each basic graph pattern to get three embeddings $\entemb_{1?}$, $\entemb_{2?}$, and $\entemb_{3?}$. $\inter()$ (red arrows) is used later on to integrate them into one single embedding $\entemb_{?}$ or $\queryembed$ for the target variable \texttt{?Place}.

In this work, we follow the same query embedding computation process. Furthermore, we extent the current model architecture to do spatial semantic lifting. \section{$\spexkge$ Model} \label{sec:method}

Since many geographic questions highly rely on spatial information (e.g., coordinates) and spatial reasoning, a spatially-explicit model is desired for the geographic logic query answering task. Moreover, the spatial semantic lifting task (Task \ref{def:spasemlift}) is only possible if we have an entity encoder which can encode the spatial information of geographic entities as well as a specially designed projection operator. 
To solve these problem, we propose a new entity encoder $\enc()$ (See Section \ref{subsec:entemb}) and a new projection operator (See Section \ref{subsec:proj}) for our $\spexkge$ model. Next, Task \ref{def:gqa} and \ref{def:spasemlift} require different training processes which will be discussed in Section \ref{subsec:qatask_model} and \ref{subsec:spasemlift_model}. 
$\spexkge$ extends the general logic query answering framework of  $\gqe$ \citep{hamilton2018embedding} and $\cga$ \citep{mai2019contextual} with explicit spatial embedding representations.

\subsection{Entity Encoder}  \label{subsec:entemb}
\begin{definition}[Entity Encoder: $\enc()$]
    \label{def:entenc}
    Given a geographic knowledge graph $\kg$, entity encoder $\enc() : \entset \to \Real^\embdim$ is defined as a function parameterized by $\theta_{\enc}$, which maps any entity $\ent_{i} \in \entset$ to a vector representation of $\embdim$ dimension, so called entity embedding $\entemb_{i} \in \Real^{\embdim}$. $\enc()$ consists of two parts -- the entity feature encoder $\featenc(): \entset \to \Real^{\fedim}$ and the entity space encoder $\spaenc(): \entset \to \Real^{\pedim}$. These two encoders map any entity $\ent_{i} \in \entset$ to a feature embedding $\feemb_{i} \in \Real^{\fedim}$ and space embedding $\peemb_{i} \in \Real^{\pedim}$,
    respectively. The final entity embedding $\entemb_{i}$ is the concatenation of $\feemb_{i}$ and $\peemb_{i}$, i.e., :
    \begin{equation}
        \entemb_{i} = \enc(\ent_{i}) = [\featenc(\ent_{i});\spaenc(\ent_{i})] = [\feemb_{i};\peemb_{i}]
    \label{equ:entemb}
    \end{equation}
\end{definition}

Here $[;]$ denotes vector concatenation of two column vectors and $\embdim = \fedim + \pedim$. $\featenc()$ has been defined in Definition \ref{def:feenc}.

\subsubsection{Entity Space Encoder}
In our work, and instead of calling them \textit{location encoder} and \textit{location embedding} \citep{mac2019presence},  we use the term  \textit{space encoder} to refer to the neural network model that encodes the spatial information of an entity and call the encoding results \textit{space embeddings}. While location encoder focus on encoding one single point location, our space encoder $\spaenc()$ aims at handling spatial information of geographic entities at different scales:
\begin{enumerate}
    \item For a small geographic entity $\ent_{i} \in \ptentset \setminus \pgonentset$ such as radio stations or restaurants, we use its location $\bx_i = \ptfun(\ent_{i})$ as the input to $\spaenc()$.
    \item For an geographic entity with a large extent $\ent_{i} \in \pgonentset$ such as countries and states, at each encoding time, we randomly generate a point $\bx_{i}^{(t)}$ as the input for $\spaenc()$ based on the 2D uniform distribution defined on its spatial extent (bounding box) $\pgonfun(\ent_{i}) = [\bx_{i}^{min};\bx_{i}^{max}]$, i.e., $\bx_{i}^{(t)} \sim \uniondist(\bx_{i}^{min},\bx_{i}^{max})$. Since during training $\spaenc()$ will be called multiple times, it will at the end learn a uniform distribution over $\ent_{i}$'s bounding box. In practice, one can sample using any process, such as stratified random sampling, or vary the sampling density by expected variation.

    \item For non-geographic entity $\ent_{i} \in \entset \setminus \ptentset$, we randomly initialize its space embedding. One benefit of this approach is that during the KG embedding training process, these embeddings will be updated based on back propagation in neural networks so that the spatial information of its connected entities in $\kg$ will propagate to this embedding as its pseudo space footprint. 
    For example, a person's spatial embedding will be close to the embedding of his/her birthplace or hometown. 
\end{enumerate}

\begin{figure}[t!] 
	\centering
	\includegraphics[width=1.0\columnwidth]{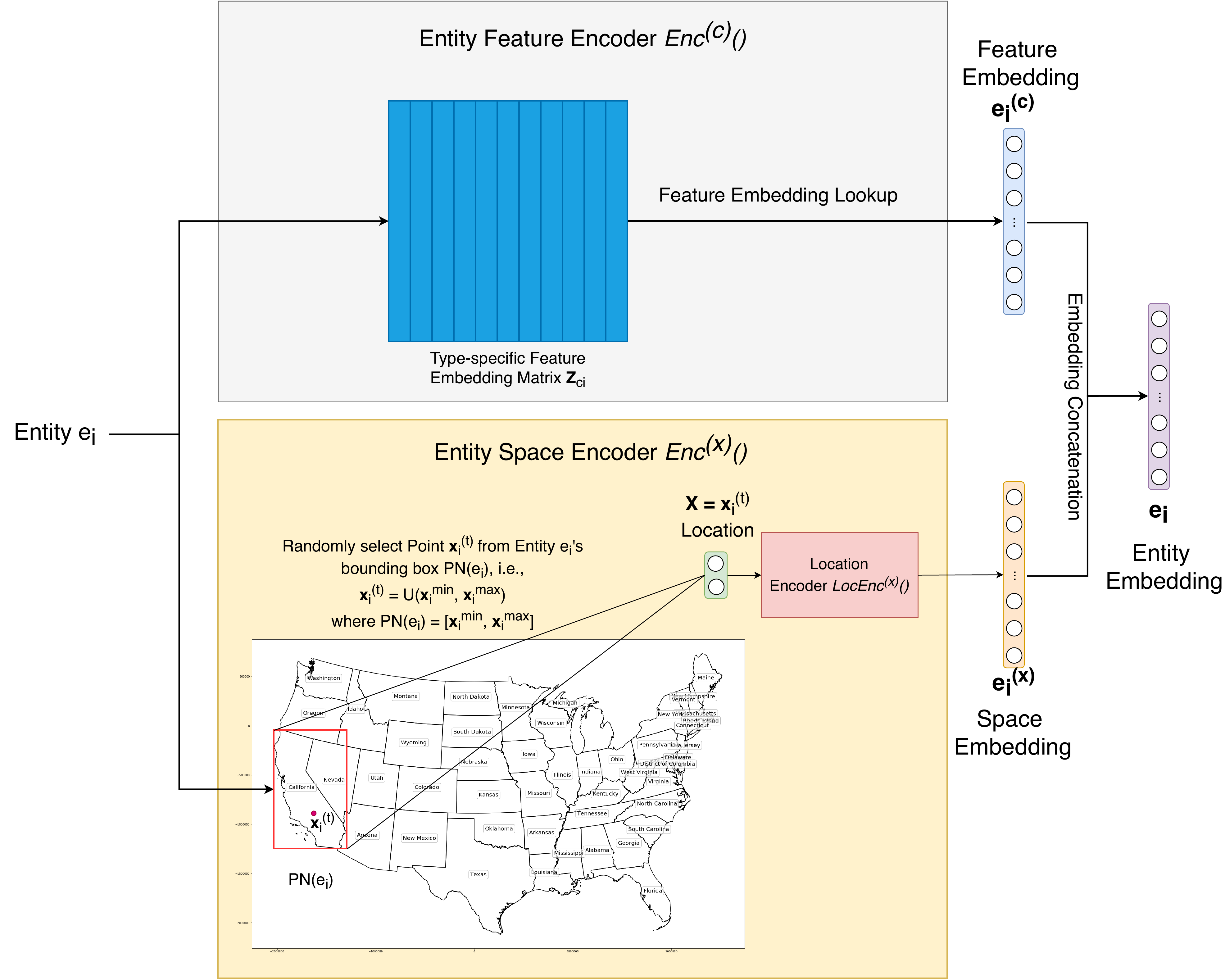}
	\caption{
	The entity encoder $\enc()$ of $\spexkge$. Compared with previous work (Figure \ref{fig:entenc_gqe}) an entity space encoder component $\spaenc()$ is added to capture the spatial information of geographic entities. }
	\label{fig:entenc}
\end{figure}

The entity space encoder $\spaenc()$ is formally defined as follow:
\begin{definition}[Entity Space Encoder: $\spaenc()$]
    \label{def:peenc}
    Given any entity $\ent_{i} \in \entset$ from $\kg$, $\spaenc()$ computes the space embedding $\peemb_{i} = \spaenc(\ent_i) \in \Real^{\pedim}$ by
    \begin{equation}
\peemb_{i} = \begin{cases}
    \locenc(\bx_i)  \; , where \;  \bx_i = \ptfun(\ent_{i})  \; , & if \; \ent_{i} \in \ptentset \setminus \pgonentset  \\
    \locenc(\bx_{i}^{(t)})   \; , where \;  \bx_{i}^{(t)} \sim \uniondist(\bx_{i}^{min},\bx_{i}^{max}) 
\; ,  \;  \pgonfun(\ent_{i}) = [\bx_{i}^{min};\bx_{i}^{max}]  
    \; , & if \; \ent_{i} \in \pgonentset  \\
    \dfrac{\embmat_{x}\onehotvec_{i}^{(x)}}{\parallel\embmat_{x}\onehotvec_{i}^{(x)}\parallel_{L2}} ,  \; &  if \; \ent_{i} \in \entset \setminus \ptentset
\end{cases}
    	\label{equ:peemb}
    \end{equation}
    
\end{definition}
Here $\embmat_{x}$ and $\onehotvec_{i}^{(x)}$ are the embedding matrix  and one-hot vector for non-geographic entities in entity space encoder $\spaenc()$ similar to Equation \ref{equ:feemb}.  $\locenc()$ denotes a location encoder module (See Equation \ref{equ:locenc}). Figure \ref{fig:entenc} illustrates the architecture of entity encoder $\enc()$. Compared with GQE's entity encoder $\encgqe()$ shown in Figure \ref{fig:entenc_gqe}, the proposed entity encoder of $\spexkge$ adds the entity space encoder $\spaenc()$ which leverages a multi-scale grid cell representation to capture the spatial information of geographic entities.

As far as using a bounding box as approximation is concerned, one reason to use bounding boxes instead of the real geometries is that doing point-in-polygon operation in real time during ML model training is very expensive and not efficient. Many spatial databases use bounding boxes as approximations of the real geometries to avoid intensive computation. We adopt the same strategy here. Moreover, the detailed spatial footprint of $\ent_i$ is expected to be captured through the training process of the entity embedding. For example, even if the model is only aware of the bounding box of California, by using the \texttt{dbo:isPartOf} relations between California and its subdivisions, the model will be informed of all the spatial extents of its subdivisions.

 \subsection{Projection Operator}  \label{subsec:proj}

\begin{definition}[Projection Operator $\proj()$]
    \label{def:proj}
    Given a geographic knowledge graph $\kg$, a projection operator $\proj() : \entset \cup \studyarea \times \relset \to \Real^\embdim$ maps a pair of $(\ent_i, \rel)$, $(\variable_{i}, \rel)$, or $(\bx_i, \rel)$, to an 
    embedding $\entemb_{i}^{\prime}$. According to the input, $\proj()$ can be treated as: 
    \textbf{(1) link prediction $\projent(\ent_{i}, \rel)$:} given a triple's head entity $\ent_i$ and relation $\rel$, predicting the tail; (2) \textbf{link prediction $\projent(\variable_{i}, \rel)$:} given a basic graph pattern $\gp = \rel(\variable_{i}, \variable_{j})$ and $\varemb_{i}$ which is the computed embedding for the existentially quantified bound variable $\variable_{i}$, predicting the embedding for Variable $\variable_{j}$;
    \textbf{(2) spatial semantic lifting $\projloc(\bx_{i}, \rel)$:} given an arbitrary location $\bx_i$ and relation $\rel$, predicting the most probable linked entity.
    Formally, $\proj()$ is defined as:
	
    \begin{equation}
    	\entemb_{i}^{\prime} = \begin{cases}
    	\projent(\ent_{i}, \rel) = \blockdiag(\projfeatmat, \projspamat)\enc(\ent_{i}) = \blockdiag(\projfeatmat, \projspamat)\entemb_{i} \; & if \; input \; =  (\ent_{i}, \rel)    \\
    	\projent(\variable_{i}, \rel) = \blockdiag(\projfeatmat, \projspamat)\varemb_{i} \; & if \; input \; =  (\variable_{i}, \rel)    \\
    	\projloc(\bx_{i}, \rel) = \blockdiag(\projspatofeatmat, \projspamat)[\locenc(\bx_i);\locenc(\bx_i)] \; & if \; input \; =  (\bx_{i}, \rel)
    	\end{cases}
    	\label{equ:proj}
	\end{equation}
	
	where $\projfeatmat \in \Real^{\fedim \times \fedim}$, $\projspamat \in \Real^{\pedim \times \pedim}$, and $\projspatofeatmat \in \Real^{\fedim \times \pedim}$ are three trainable and relation-specific matrices. $\projfeatmat$ and $\projspamat$ focus on the feature embedding and space embedding. $\projspatofeatmat$ transforms the space embedding 
    $\peemb_{i}$ to its correspondence in feature embedding space. $\blockdiag(\projfeatmat, \projspamat) \in \Real^{\embdim \times \embdim} $ and $\blockdiag(\projspatofeatmat, \projspamat) \in \Real^{\embdim \times 2\pedim} $ indicate two block diagonal matrices based on $\projfeatmat$, $\projspamat$, and $\projspatofeatmat$. $[\locenc(\bx_i);\locenc(\bx_i)]$ indicates the concatenation of two identical space embedding $\locenc(\bx_i)$. Here, we use the same $\projent()$ for the first two cases to indicate they share the same neural network architecture. This is because both of them are link prediction tasks with different inputs.
\end{definition}

\begin{figure}[h!]
	\centering
	\includegraphics[width=1.0\columnwidth]{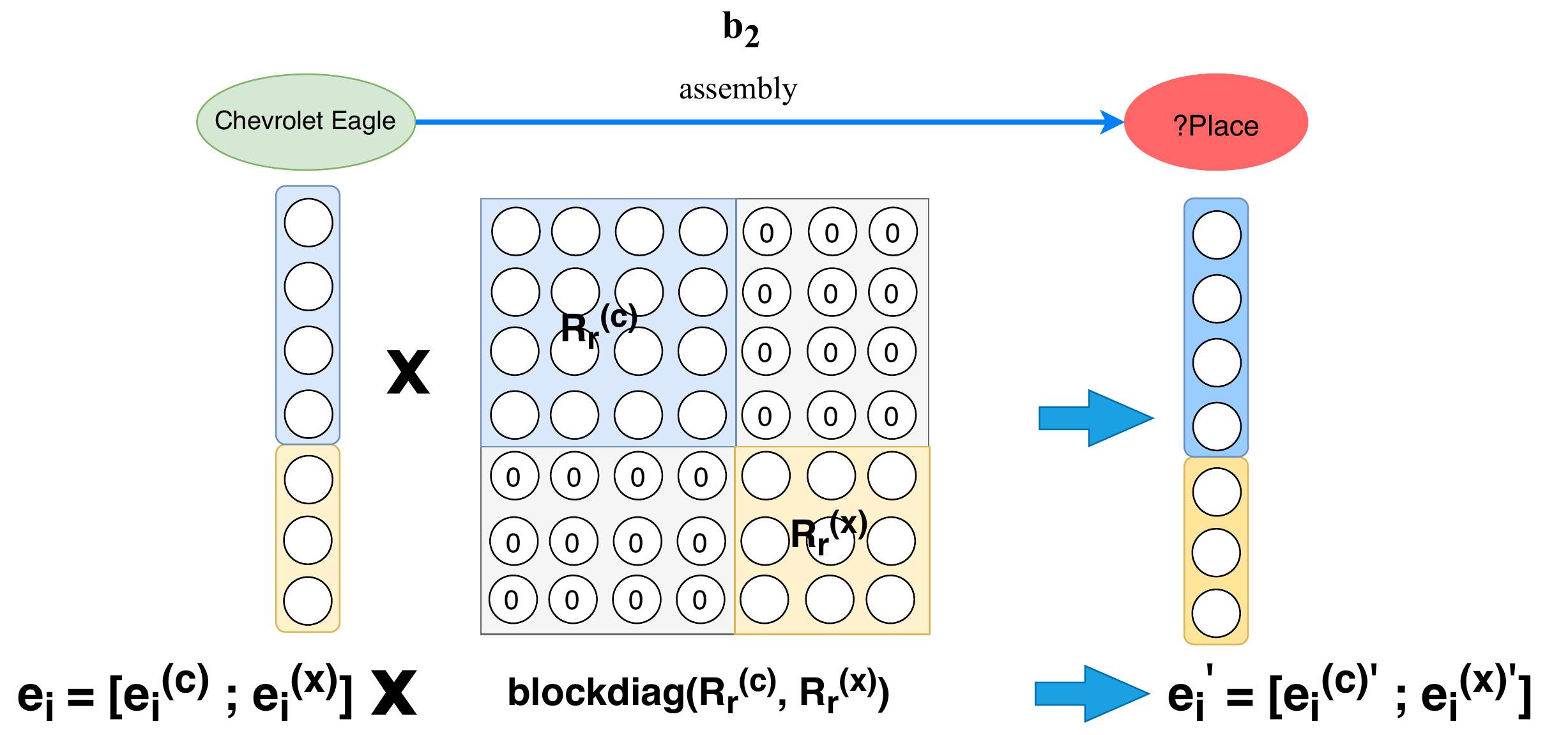}
	\caption{An illustration of projection operator $\projent()$ of $\spexkge$ with the input $(\ent_{i}, \rel)$.} 
	\label{fig:proj}
\end{figure}

\textbf{Link Prediction}: Figure \ref{fig:proj} illustrates the idea of projection operator $\projent()$ by using the basic graph pattern $\bgptwo$ in $\queryC$ (See Figure \ref{fig:qg}) as an example (the first case). 
Given the embedding of \texttt{dbr:Chevrolet\_Eagle} and the relation-specific matrix $\blockdiag(\projfeatmat, \projspamat)$ for relation \texttt{dbo:assembly}, we can predict the embedding of the variable \texttt{?Place} - $\entemb_{?2}$. 
 
\begin{figure}[h!]
	\centering
	\includegraphics[width=1.0\columnwidth]{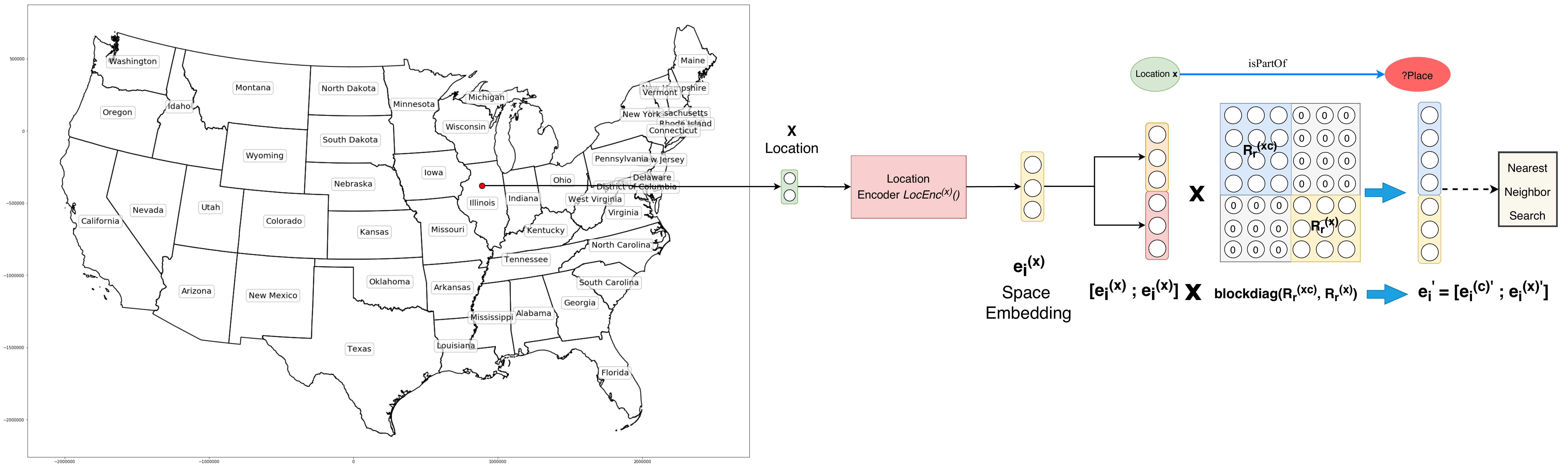}
	\caption{Spatial semantic lifting in the embedding space by using $\enc()$ and $\projloc()$} 
	\label{fig:spasemlift}
\end{figure}

\textbf{Spatial Semantic Lifting}: Figure \ref{fig:spasemlift} shows how to use $\projloc()$ in the semantic lifting task. See Section \ref{subsec:spasemlift_model} for detail description. Note that ``$\times$'' in Figure \ref{fig:proj} and \ref{fig:spasemlift} indicates $\blockdiag(\projfeatmat, \projspamat)\entemb_{i}$ and $\blockdiag(\projspatofeatmat, \projspamat)[\locenc(\bx_i);\locenc(\bx_i)]$.

\subsection{Geographic Logic Query Answering $\kgqefun_{\kg,\theta}(\cgq)$ Model Training}  \label{subsec:qatask_model}

We train the $\spexkge$ on both the original knowledge graph structure with an unsupervised objective $\kgtloss$ and the query-answer pairs with a supervised objective $\qatloss$ (See Equation \ref{equ:loss}):

\begin{equation}
\loss^{(QA)} = \kgtloss + \qatloss
\label{equ:loss}
\end{equation}

\paragraph{Unsupervised KG Training Phase} \label{pg:kgtrain}
In this phase, we train $\spexkge$ components  based on the local KG structure. In $\kg = \kgdefs$, for every entity $\ent_{i} \in \entset$, we first obtain its 1-degree neighborhood $\nei(\ent_{i}) = \{(\rel_{ui}, \ent_{ui}) | \rel_{ui}(\ent_{ui}, \ent_{i}) \in \kg\} \cup \{(\rel_{oi}^{-1}, \ent_{oi}) | \rel_{oi}(\ent_{i}, \ent_{oi}) \in \kg\}$. We sample $\neisize$ tuples from $\nei(\ent_{i})$ to form a sampled neighborhood $\neisamp(\ent_{i}) \subseteq \nei(\ent_{i})$ and $|\neisamp(\ent_{i})| = \neisize$. We treat this subgraph as a conjunctive graph query with $\neisize$ basic graph patterns, in which entity $\ent_{i}$ holds the target variable position. 
The model predicts the embedding of $\ent_i$ such that the correct embedding $\entemb_i$ is the closest one to the predicted embedding $\entemb_{i}^{\prime\prime}$ against all embeddings $\entemb_{i}^{-}$ in negative sample set $\negsamp(\ent_{i})$:
\begin{equation}
\kgtloss = \sum_{\ent_{i} \in \entset} \sum_{\ent_{i}^{-} \in \negsamp(\ent_{i})} max(0, \Delta - \cosine(\kgtembed(\ent_{i}), \entemb_{i}) + \cosine(\kgtembed(\ent_{i}), \entemb_{i}^{-}))
\label{equ:kgloss}
\end{equation}

where
\begin{equation}
\entemb_{i}^{\prime\prime} = \kgtembed(\ent_{i}) = \inter(\{\projent(\ent_{ci}, \rel_{ci}) | (\rel_{ci}, \ent_{ci}) \in \neisamp(\ent_{i})\})
\label{equ:kgtrain}
\end{equation}

Here $\kgtloss$ is a max-margin loss and $\Delta$ is the margin.

\begin{figure}[h!]
	\centering
	\includegraphics[width=0.8\columnwidth]{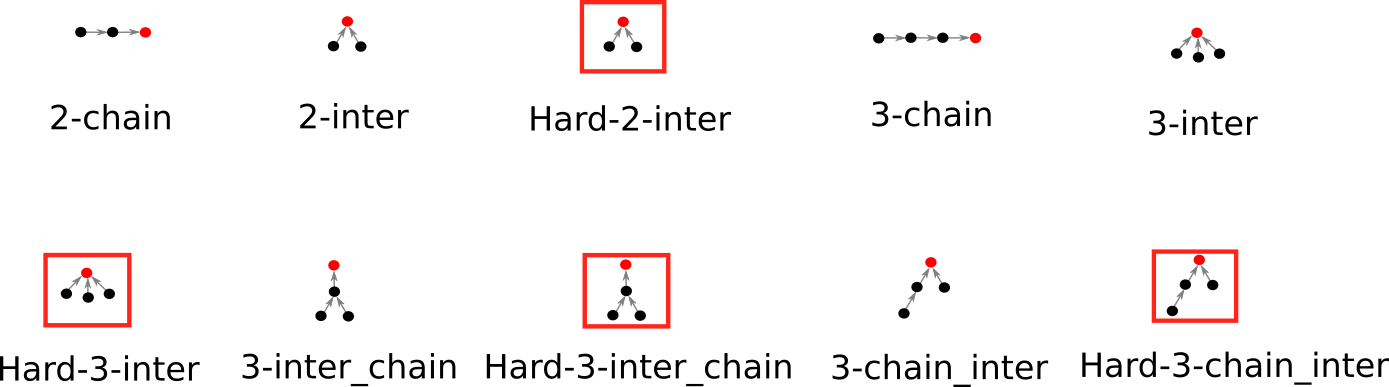}
	\caption{The DAG structures of the conjunctive graph queries we sampled from $\kg$. Nodes indicates entities or variables and edges indicate basic graph patterns. The red node is the target variable of the corresponding query. the DAG structures surrounded by red boxes indicate queries sampled with hard negative sampling method.} 
	\label{fig:dag}
\end{figure}

\paragraph{Supervised Query-Answer Pair Training Phase} \label{pg:qatrain}

We train $\spexkge$ by using conjunctive query-answer pairs. We first sample $\qasamplesize$\ different conjunctive graph query (logical query)-answer pairs $\qaset = \{(\query_{i}, \answer_{i})\}$ from $\kg$.
We treat each entity as the target variable of a CQG and sample $K$ queries for each DAG structure. All DAG structures we considered in this work are shown in Figure \ref{fig:dag}. The way to do query sampling is to sort the nodes in a DAG in a topological order and sample one basic graph pattern at one time by following this order and navigating on the $\kg$ \citep{hamilton2018embedding}. \textbf{In order to generate geographic conjunctive graph query, we have the restriction $\ent_i \in \ptentset$.}

The training objective is to make the correct answer entity embedding $\answerembed_{i}$ be the closest one to the predicted query embedding $\queryembed_{i} = \kgqefun_{\kg,\theta}(\cgq_i)$ 
against all the negative answers' embeddings $\answerembed_{i}^{-}$ in negative answer set $\negsamp(\query_{i}, \answer_{i})$. We also use a max-margin loss:
\begin{equation}
\qatloss = \sum_{(\cgq_{i}, \answer_{i}) \in \qaset} \sum_{\answer_{i}^{-} \in \negsamp(\cgq_{i}, \answer_{i})} max(0, \Delta - \cosine(\queryembed_{i}, \answerembed_{i}) + \cosine(\queryembed_{i}, \answerembed_{i}^{-}))
\label{equ:qaloss}
\end{equation}
For $\negsamp(\cgq_{i}, \answer_{i})$ we compared two negative sampling strategies : 1) \textit{negative sampling}: $\negsamp(\query_{i}, \answer_{i}) \subseteq \entset$ is a fixed-size set of entities such that $\forall \ent_{i}^{-} \in \negsamp(\query_{i}, \answer_{i}), \; \typefun(\ent_{i}^{-}) = \typefun(\ent_{i})  \; and \; \ent_{i}^{-} \neq \ent_{i}$; 2) \textit{hard negative sampling}: $\negsamp(\cgq_{i}, \answer_{i})$ is a fixed-size set of entities which satisfy some of the basic graph patterns $b_{ij}$ (See Definition \ref{def:cgq}) in $\cgq_i$ but not all of them.

\subsection{Spatial Semantic Lifting $\spasemliftfun_{\kg,\theta_{ssl}}(\bx, \rel)$ Model Training}  \label{subsec:spasemlift_model}

We randomly select a point  $\bx_i \in \studyarea \subseteq \Real^{2}$ from the study area, and use location encoder $\locenc$ to encode its location embedding $\peemb_{i} \in \Real^{\pedim}$. 
Since we do not have the feature embedding for this location, to make the whole model as an inductive learning one, we use $\projloc()$
to predict the tail embedding $\spasemliftentemb = \spasemliftfun_{\kg,\theta_{ssl}}(\bx_i, \rel) $ of this virtual triple $\rel(\bx_i, \spasemliftent)$. This is equivalent to ask a query $\rel(\bx_i, ?\ent)$ to $\kg$. A nearest neighbor search in the entity embedding space will produce the predicted entity who can link to location $\bx_i$ with relation $\rel$. 
Since given any location $\bx_i$ from the study area, $\spasemliftfun_{\kg,\theta_{ssl}}(\bx_i, \rel)$ can predict the entity embedding that $\bx_i$ can link to given relation $\rel$, this is a fully inductive learning based model. This model does not require location $\bx_i$ to be selected from a predefined set of locations which is a requirement for transductive learning based models such as \citet{kejriwal2017neural}.
Figure \ref{fig:spasemlift} shows the idea of spatial semantic lifting.

We train the spatial semantic lifting model $\spexkge_{ssl}$ with $\enc()$, $\projent()$, and $\projloc()$ by using two objectives: link prediction objective $\lploss$ and spatial semantic lifting objective $\sslloss$.

\begin{equation}
\loss^{(SSL)} = \lploss + \sslloss
\label{equ:ssl_loss}
\end{equation}

\paragraph{Link Prediction Training Phase}  \label{pg:lptrain}
The link prediction training phase aims at training the feature embeddings of each entity. For each triple $\triple_{i} = (h_{i},\rel_{i},t_{i}) \in \triset$, we can use $\enc()$ and $\projent()$ to predict the tail entity embedding given the head and relation - $\projent(h_{i},\rel_{i})$ - or predict the head entity embedding given the tail and relation - $\projent(t_{i},\rel_{i}^{-1})$. Note that we have two separate $\projent()$ for $\rel_i$ and $\rel_i^{-1}$. Equation \ref{equ:lploss} shows the loss function where $\negsamp_{t}(\ent_{i})$ is the set of negative entities who share the same type with entity $\ent_{i}$.

\begin{align}
\lploss = \sum_{\triple_{i} = (h_{i},\rel_{i},t_{i}) \in \triset} \sum_{t_{i}^{-} \in \negsamp_{t}(t_{i})} max(0, \Delta - \cosine(\projent(h_{i},\rel_{i}), \mathbf{t}_{i}) + \cosine(\projent(h_{i},\rel_{i}), \mathbf{t}_{i}^{-}))  \nonumber \\
+  \sum_{\triple_{i} = (h_{i},\rel_{i},t_{i}) \in \triset} \sum_{h_{i}^{-} \in \negsamp_{t}(h_{i})} max(0, \Delta - \cosine(\projent(t_{i},\rel_{i}^{-1}), \mathbf{h}_{i}) + \cosine(\projent(t_{i},\rel_{i}^{-1}), \mathbf{h}_{i}^{-}))
\label{equ:lploss}
\end{align}

\paragraph{Spatial Semantic Lifting Training Phase}  \label{pg:ssltrain}
We also directly optimize our model on the spatial semantic lifting objective. We denote $\triset_{s}$ and $\triset_{o}$ as sets of triples whose head (or tail) entities are geographic entities, i.e., $\triset_{s} = \{\triple_{i} |  \triple_{i} = (h_{i},\rel_{i},t_{i}) \in \triset \land h_{i} \in \ptentset \}$
and 
$\triset_{o} = \{\triple_{i} | \triple_{i} = (h_{i},\rel_{i},t_{i}) \in \triset \land t_{i} \in \ptentset \}$.  The training objective is to make the tail entity embedding $\mathbf{t}_{i}$ to be the closest one to the predicted embedding $\projloc(\geofun(h_{i}),\rel_{i})$ 
against all negative entity embeddings $\mathbf{t}_{i}^{-}$. We do the same for the inverse triple $(t_{i},\rel_{i}^{-1},h_{i})$. The loss function is shown in Equation \ref{equ:sslloss}.

\begin{align}
\sslloss = \sum_{\triple_{i} = (h_{i},\rel_{i},t_{i}) \in \triset_{s}} \sum_{t_{i}^{-} \in \negsamp_{t}(t_{i})} max(0, \Delta - \cosine(\projloc(\geofun(h_{i}),\rel_{i}), \mathbf{t}_{i}) + \cosine(\projloc(\geofun(h_{i}),\rel_{i}), \mathbf{t}_{i}^{-}))  \nonumber \\
+  \sum_{\triple_{i} = (h_{i},\rel_{i},t_{i}) \in \triset_{o}} \sum_{h_{i}^{-} \in \negsamp_{t}(h_{i})} max(0, \Delta - \cosine(\projloc(\geofun(t_{i}),\rel_{i}^{-1}), \mathbf{h}_{i}) + \cosine(\projloc(\geofun(t_{i}),\rel_{i}^{-1}), \mathbf{h}_{i}^{-}))
\label{equ:sslloss}
\end{align}

where

\begin{equation}
\geofun(\ent_{i}) = \begin{cases}
\bx_i = \ptfun(\ent_{i})  \; , & if \; \ent_{i} \in \ptentset \setminus \pgonentset  \\
\bx_{i}^{(t)} \sim \uniondist(\bx_{i}^{min},\bx_{i}^{max}) 
\; ,  \;  \pgonfun(\ent_{i}) = [\bx_{i}^{min};\bx_{i}^{max}]  
\; , & if \; \ent_{i} \in \pgonentset  
\end{cases}
\label{equ:ent2x}
\end{equation} \section{Experiment} \label{sec:exp}

To demonstrate how $\spexkge$ incorporates spatial information of geographic entities such as locations and spatial extents we experimented with two tasks --  
geographic logic query answering and spatial semantic lifting. 
To demonstrates  the effectiveness of spatially explicit models and the importance to considering the scale effect in location encoding we select multiple baselines on the geographic logic query answering task. 
To show that $\spexkge$ is able to  link a randomly selected location to entities in the existing KG with some relation, which 
none of the existing KG embedding models can solve, we proposed a new task - spatial semantic lifting.

\subsection{$\dbgeo$ Dataset Generation} \label{subsec:data}
In order to evaluate our proposed location-aware knowledge graph embedding model $\spexkge$, we first build a geographic knowledge graph which is a subgraph of DBpedia by following the common practice in KG embedding research \citep{bordes2013translating,wang2014knowledge,mai2019relaxing}. We select the mainland of United States as the study area $\studyarea$ since previous research \citep{janowicz2016moon} has shown that DBpedia has relatively richer geographic coverage in United States.
The KG construction process is as follows:

\begin{enumerate}
    \item We collect all the geographic entities within the mainland of United States as the seed entity set $\entset_{seed}$ which accounts for 18,780 geographic entities\footnote{We treat an entity as a geographic entity if its has a \texttt{geo:geometry} triple in DBpedia};
    We then collect their 1- and 2-degree object property triples with \texttt{dbo:} prefix predicates/relations\footnote{\url{http://dbpedia.org/sparql?help=nsdecl}}; 
    \item We compute the degree of each entity in the collected KG and delete any entity, together with its corresponding triples, if its node degree is less than a threshold $\datathreshold$. We use $\datathreshold = 10$ for non-geographic entities and $\datathreshold = 5$ for geographic entities, because many geographic entities, such as radio stations, have fewer object type property triples and a smaller threshold ensures that a relative large number of geographic entities can be extracted from the KG;
    \item We further filter out those geographic entities that are newly added from Step 2 and are outside of the mainland of United States. The resulting triples form our KG, and we denote the geographic entity set as $\ptentset$.
    \item We split $\kg$ into training, validation, and testing triples with a radio of 90:1:9 so that every entity and relation appear in the training set. We denote the knowledge graph formed by the training triples as $\kg_{train}$ while denoting the whole KG as $\kg$.
    \item We generate $K$ conjunctive graph query-answer pairs from $\kg$ for each DAG structure shown in Figure \ref{fig:dag} based on the query-answer generation process we described in Section \ref{subsec:qatask_model}. $Q(\kg)$ and $Q(\kg)_{geo}$ indicate the resulting QA set while $Q_{geo}(\kg)$ indicates the geographic QA set. For each query $\cgq_i$ in training QA set, we make sure that each query is answerable based on $\kg_{train}$, i.e., $\qagoldfun(\kg_{train}, \cgq_i) \neq \emptyset$. As for query $\cgq_i$ in validation and testing QA set, we make sure each query $\cgq_i$ satisfies $\qagoldfun(\kg_{train}, \cgq_i) = \emptyset$ and $\qagoldfun(\kg, \cgq_i) \neq \emptyset$.
    \item For each geographic entity $\ent \in \ptentset$, we obtain its location/coordinates by extracting its \texttt{geo:geometry} triple from DBpedia. We project the locations of geographic entities into US National Atlas Equal Area projection coordinate system (epsg:2163) $\coordsys$. $\ptfun(\ent) = \bx$ indicates the location of $\ent$ in the projection coordinate system $\coordsys$. 
    \item For each geographic entity $\ent \in \ptentset$, we get its spatial extent (bounding box) $\pgonfun(\ent)$ in $\coordsys$ by using ArcGIS Geocoding API\footnote{\url{https://geocode.arcgis.com/arcgis/rest/services/World/GeocodeServer/find}} and OpenStreetMap API. 80.6\% of geographic entities are obtained. We denote them as $\pgonentset$.
    \item For each entity $\ent_i \in \entset$, we obtain its types by using \texttt{rdf:type} triples. Note that there are entities having multiple types. We look up the DBpedia Ontology (class hierachy) to get their level-1 superclass. We find out that every entity in $\kg$ has only one level-1 superclass type. Table \ref{tab:type} shows statistics of entities in different types.
    \item To build the training/validation/testing datasets for spatial semantic lifting, we obtain $\triset_{s}, \triset_{o} \subseteq \triset$ (See Section \ref{subsec:spasemlift_model}), each triple of which is composed of geographic entities as its head or tail. We denote $\relset_{ssl} = \{\rel_{i} | \triple_{i} = (h_{i},\rel_{i},t_{i}) \in \triset_{s} \cap \triset_{o} \}$
    
\end{enumerate}

\begin{figure}[h!]
	\centering
	\includegraphics[width=1.0\columnwidth]{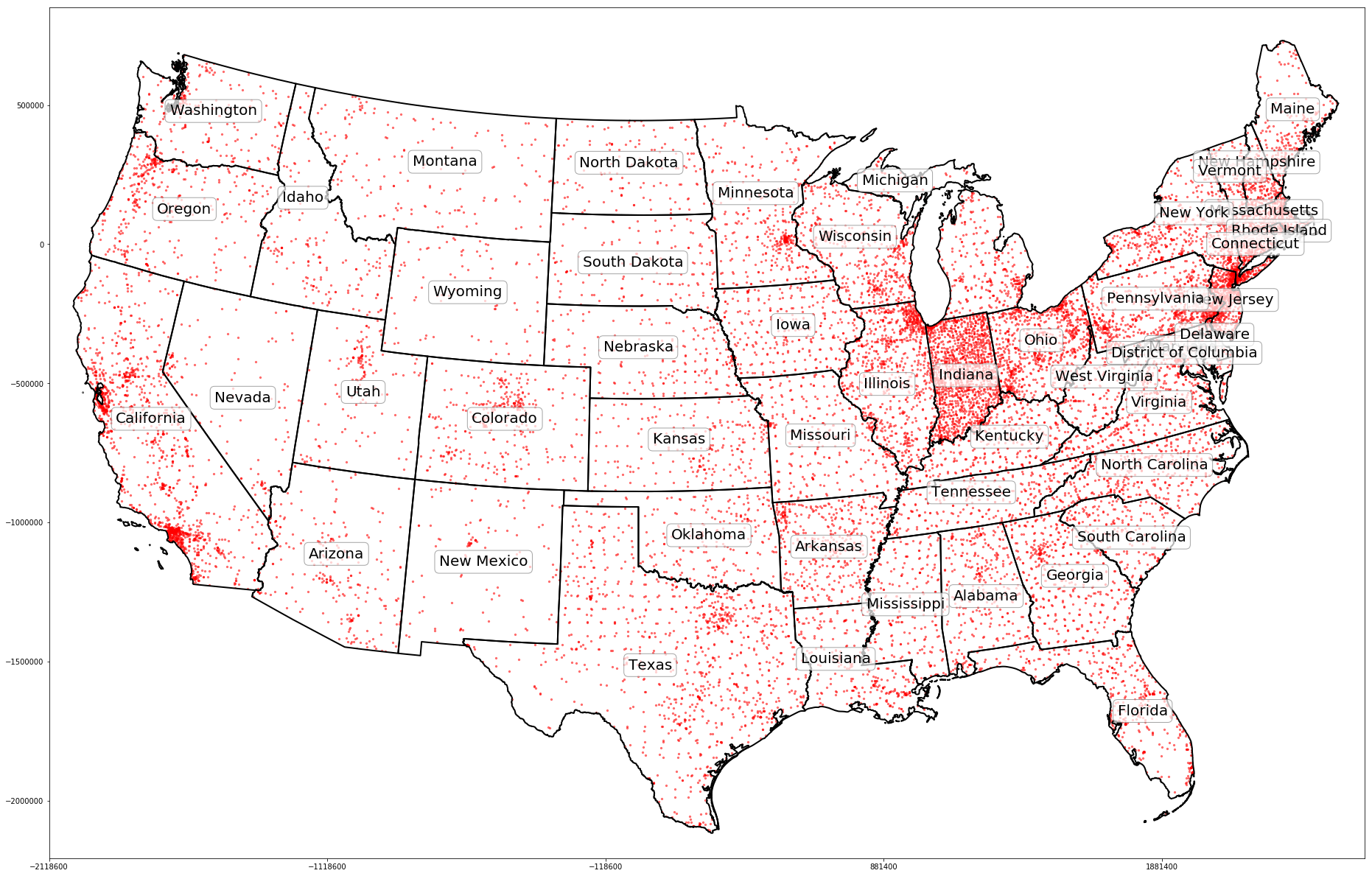}
	\caption{Spatial distribution of all geographic entities in $\kg$} 
	\label{fig:entmap}
\end{figure}

We denote $Q^{(2)}(\kg)$, $Q^{(3)}(\kg)$ as the general QA sets which contain 2 and 3 basic graph patterns, and similarly for $Q_{geo}^{(2)}(\kg)$, $Q_{geo}^{(3)}(\kg)$. Table \ref{tab:data} shows the statistics of the constructed $\kg$, the generated QA sets, and the spatial semantic lifting dataset in $\dbgeo$.
Figure \ref{fig:entmap} shows the spatial distribution of all geographic entities $\ptentset$ in $\kg$.

\begin{table*}
	\caption{Statistics for 
	our	dataset in \dbgeo~ (Section~\ref{subsec:data}).
	``XXXX/QT'' indicates the number of QA pairs per query type.}
	\label{tab:data}
	\centering
	
\begin{tabular}{c|c|ccc}
\toprule
&                                         & \multicolumn{3}{c}{\dbgeo}        \\
&                                         & Training & Validation & Testing  \\ \hline
\multirow{5}{*}{Knowledge Graph} & $|\triset|$                              & 214,064    & 2,378      & 21,406    \\
& $|\relset|$                              & 318        & -          & -        \\
& $|\entset|$                              & 25,980     & -          & -        \\
& $|\ptentset|$                            & 18,323     & -          & -        \\
& $|\pgonentset|$                          & 14,769     & -          & -        \\ \hline
\multirow{4}{*}{Geographic Question Answering} & $|Q^{(2)}(\kg)|$                         & 1,000,000  & -          & -        \\
& $|Q^{(3)}(\kg)|$                         & 1,000,000  & -          & -        \\
& $|Q_{geo}^{(2)}(\kg)|$                   & 1,000,000  & 1000/QT    & 10000/QT \\
& $|Q_{geo}^{(3)}(\kg)|$                   & 1,000,000  & 1000/QT    & 10000/QT \\ \hline
\multirow{2}{*}{Spatial Semantic Lifting} & $|\triset_{s} \cap \triset_{o}|$         & 138,193    & 1,884      & 17,152    \\
& $|\relset_{ssl}|$                        & 227        & 71         & 135    \\
\bottomrule
\end{tabular}
\vspace{-0.3cm}
\end{table*}

\begin{table}
	
	\caption{Number of entities for each entity type in $\dbgeo$}
	\label{tab:type}
	\centering
\begin{tabular}{l|r}
\toprule
Entity Type              & Number of Entities    \\ \hline
dbo:Place                & 16,527 \\
dbo:Agent                & 8,371  \\
dbo:Work                 & 594   \\
dbo:Thing                & 179   \\
dbo:TopicalConcept       & 134   \\
dbo:MeanOfTransportation & 104   \\
dbo:Event                & 71    \\
\bottomrule
\end{tabular}
\end{table}

\subsection{Evaluation on the Geographic Logic Query Answering Task} \label{subsec:qa_eval}

\subsubsection{Baselines}  \label{subsubsec:qabaseline}
In order to quantitatively evaluate 
$\spexkge$ on geographic QA task, we train $\spexkge_{full}$ and multiple baselines on $\kg$ in $\dbgeo$. Compared to previous work \citep{hamilton2018embedding,mai2019contextual}, the most important contribution of this work is the entity space encoder $\spaenc()$ which makes our model spatially explicit. So we carefully select four baselines to test the contribution of $\spaenc()$ on the geographic logic QA task. 
We have selected four baselines:
\begin{enumerate}
	\item $\gqediag$ and $\gqe$: two versions of the logic query answering model proposed by \citet{hamilton2018embedding} which have been discussed in detail in Section \ref{sec:qa_background}. The main different between $\gqediag$ and $\gqe$ is the projection operator they use: $\projgqediag$ and $\projgqe$ accordingly. Compared with $\spexkge_{full}$, both $\gqediag$ and $\gqe$ only use entity feature encoder $\featenc()$ as the entity encoder and $\intergqe$ as the intersection operator. Both methods only use $\qatloss$ in Equation \ref{equ:loss} as the training objective. There two baselines are implemented based on the original code repository\footnote{\url{https://github.com/williamleif/graphqembed}} of \citet{hamilton2018embedding}.
	\item $\cga$: a logic query answering model proposed by \citet{mai2019contextual} (See Section \ref{sec:qa_background}). Compared with $\spexkge_{full}$, $\cga$ uses different entity encoder ($\enccga$) and projection operator ($\projcga$) such that the spatial information of each geographic entity is not considered.
	This baseline is used to test whether designing spatially explicit logic query answering model can outperform general models on the geographic query answering task.
	\item $\spexkge_{direct}$: a simpler version of $\spexkge_{full}$ which uses a \textbf{single scale location encoder} in the entity encoder instead of the multi-scale periodic location encoder as shown in Equation \ref{equ:locenc} in Section \ref{subsec:locenc_related}. Instead of first decomposing input $\bx$ into a multi-scale periodic representation by using sinusoidal functions with different frequencies \citep{mai2020multiscale}, the location encoder of $\spexkge_{direct}$ directly inputs $\bx$  into a feed forward network. This single-scale location encoder is proposed in \citet{mai2020multiscale} as one baseline model - $direct$. Moreover, its entity space encoder does not consider the spatial extent of each geographic entity either and just uses its coordinates to do location encoding. This baseline is used to test the effectiveness of using multi-scale periodical representation learning in our $\spexkge$ framework.
	\item $\spexkge_{pt}$: a simpler version of $\spexkge_{full}$ whose entity space encoder does not consider the spatial extents of geographic entities. The only different between $\spexkge_{pt}$ and $\spexkge_{direct}$ is that $\spexkge_{pt}$ uses $Space2Vec$ \citep{mai2020multiscale} as the location encoder while $\spexkge_{direct}$ utilizes the single scale $direct$ model as the location encoder. This baseline is used to test the necessity to consider the spatial extent of geographic entities in our $\spexkge$ framework. In other words, it uses Equation \ref{equ:peemb_pt} for its space encoder:
	\begin{equation}
	\peemb_{i} = \begin{cases}
	\locenc(\bx_i)  \; , where \;  \bx_i = \ptfun(\ent_{i})  \; , & if \; \ent_{i} \in \ptentset  \\
	\dfrac{\embmat_{x}\onehotvec_{i}^{(x)}}{\parallel\embmat_{x}\onehotvec_{i}^{(x)}\parallel_{L2}} ,  \; &  if \; \ent_{i} \in \entset \setminus \ptentset
	\end{cases}
	\label{equ:peemb_pt}
	\end{equation}
	\item $\spexkge_{space}$: a simpler version of $\spexkge_{full}$ whose entity encoder does not have the feature encoder component. This baseline is used to understand how the space encoder $\spaenc()$ captures the connectivity information of $\kg$.
\end{enumerate}

\subsubsection{Training Details} \label{subsubsec:qa_train_detail}
We train our model $\spexkge_{full}$ and six baselines on $\dbgeo$ dataset. $\gqediag$ and $\gqe$ are trained on the general QA pairs and geographic QA pairs as \citet{hamilton2018embedding} did. The other models are additionally trained on the original KG structure. 
Gird search is used for hyperparameter tuning: $\embdim = [32, 64, 128]$, $\fedim = [16, 32, 64]$, $\pedim = [16, 32, 64]$, $\nscale = [8, 16, 32, 64]$, $\lambda_{min} = [10, 50, 200, 1000]$. The best performance is obtained when $\embdim = 128$, $\fedim = 64$, $\pedim = 64$, $\nscale = 16$, $\lambda_{min} = 50$. $\lambda_{max} = 5400000$ is determined by the study area $\studyarea$. We also try different activation functions (i.e., Sigmoid, ReLU, LeakyReLU) for the full connected layers $\pemlp()$ of location encoder $\locenc()$. We find out that $\spexkge_{space}$ achieves the best performance with LeakyReLU as the activation function together with L2 normalization on the location embedding. $\spexkge_{direct}$, $\spexkge_{pt}$, and $\spexkge_{full}$ obtain the best performance with Sigmoid activation function without L2 normalization on the location embedding.
We implement all models in PyTorch and train/evaluate each model on a Ubuntu machine with 2 GeForce GTX Nvidia GPU cores, each of which has 10GB memory. The $\dbgeo$ dataset and related codes will be opensourced.

\subsubsection{Evaluation Results}  \label{subsubsec:qaeval}
We evaluate $\spexkge_{full}$ and six baselines on the validation and testing QA datasets of $\dbgeo$. 
Each model produces a cosine similarity score between the predicted query embedding $\queryembed$ and the correct answer embedding $\answerembed$ (as well as the embedding of negative answers). The objective is to rank the correct answer top 1 among itself and all negative answers given their cosine similarity to $\queryembed$. Two evaluation metrics are computed: Area Under ROC curve (AUC) and Average Percentile Rank (APR). AUC compares the correct answer with one random sampled negative answer for each query. An ROC curve is computed based on model performance on all queries and the area under this curve is obtained. As for APR, the percentile rank of the correct answer among all negative answers is obtained for each query based on the prediction of a QA model. Then APR is computed as the average of the percentile ranks of all queries. Since AUC only uses one negative sample per query while APR uses all negative samples for each query. We consider APR as a more robust evaluation metric.

Table \ref{tab:qaeval} shows the evaluation results of $\spexkge_{full}$ as well as six baselines on the validation and testing QA dataset of $\dbgeo$. We split each dataset into different categories based on their DAG structures (See Figure \ref{fig:dag}). 
Note that logic query answering is a very challenging task. As for the two works which share a similar set up as ours, \citet{hamilton2018embedding} show that their $\gqe$ model outperforms TransE baseline by 1.6\% of APR on Bio dataset. Similarly, \citet{mai2019contextual} demonstrate that their $\cga$ model outperfroms $\gqe$ model by 1.39\% and 1.65\% of APR on DB18 and WikiGeo19 dataset. In this work, we show that our $\spexkge_{full}$ outperforms the current state-of-the-art $\cga$ model by 2.17\% and 1.31\% in terms of APR on the validation and testing dataset of $\dbgeo$ respectively. We regard it as a sufficient signal to show the effective of $\spexkge_{full}$ on the geographic QA task. 
Some interesting conclusions can be drawn from Table \ref{tab:qaeval}:
\begin{enumerate}
	\item $\cga$ has a significant performance improvement over $\gqediag$ and $\gqe$ on $\dbgeo$. This result is consistent with that of \citet{mai2019contextual} which demonstrates the advantage of the self-attention mechanism in $\intercga$.
    \item The performance of $\spexkge_{direct}$ and $\cga$ are similar, which shows that a simple single-scale location encoder ($\spexkge_{direct}$) is not sufficient to capture the spatial information of geographic entities.
    
    \item $\spexkge_{full}$ performs better than $\spexkge_{pt}$ which only considers the location information of geographic entities. This illustrates that scale effect is beneficial for the geographic logic QA task.
    
    \item The performance of $\spexkge_{space}$ is the worst among all models. This indicates that it is not enough to only consider spatial information as the input features for entity encoder $\enc()$. This makes sense because each entity in $\kg$ has a lot of semantic information other than their spatial information, and only using spatial information for entity embedding learning is insufficient. However, $\spexkge_{space}$ is a fully inductive learning model which enables us to do spatial semantic lifting.
	\item Compared $\spexkge_{full}$ with $\cga$, we can see that $\spexkge_{full}$ outperforms $CGA$ for almost all DAG structures on testing dataset except ``Hard-3-chain\_inter'' (-0.58\%) while top 2 DAG structures with the largest margin are ``3-inter\_chain'' (2.15\%) and ``3-chain\_inter'' (2.08\%). On the validation dataset, $\spexkge_{full}$ gets higher $\Delta APR$ compared to $CGA$ on ``Hard-3-inter\_chain'' (7.42\%) and ``3-inter\_chain'' (6.08\%). $\gqediag$ shows the best performance on ``Hard-3-chain\_inter'' query structure. 
\end{enumerate}

In order to demonstrate how the intersection operator $\inter()$ helps to improve the model performance on the geographic QA task, we show $\spexkge_{full}$'s predicted ranking list of entities on Query $\queryC$ as well as its three basic graph patterns in Table \ref{tab:qa_rank}.  These 12 entities in this table represent the hard negative sampling set of Query $\queryC$. \texttt{dbr:Oakland,\_California} is the correct answer for Query $\queryC$. We can see that the top ranked four entities of $\bgpone$: $IsPartOf^{-1}(Alameda \; County, ?Place)$ are all subdivisions of Alameda County. The top ranked 5 entities of $\bgptwo$: $Assembly(Chevrolet \; Eagle, ?Place)$ are all assembly places of Chevrolet Eagle. Similarly, the top ranked entities of $\bgpthree$: $NearestCity(San \; Francisco \; Bay, ?Place) $ are close to San Francisco Bay. The full query $\queryC$ yield the best rank of the correct answer. This indicates that each basic graph pattern contributes to the query embedding prediction of $\spexkge_{full}$. Moreover, to compare performances of different models on Query $\queryC$, the percentile rank given by $CGA$, $\spexkge_{pt}$, and $\spexkge_{full}$ are 53.9\%, 61.5\%, and 77.0\%, respectively.

\begin{sidewaystable}[h]
	{\small
	
		\caption{The evaluation of geographic logic query answering on $\dbgeo$ (using AUC (\%) and APR (\%) as evaluation metric)}
		\label{tab:qaeval}

		\begin{tabular}{c|l|cc|cc|cc||cc|cc|cc|cc}
			\toprule
			& \multicolumn{1}{c|}{DAG Type} & \multicolumn{2}{c|}{$\gqediag$}    & \multicolumn{2}{c|}{$\gqe$} & \multicolumn{2}{c||}{$\cga$} & \multicolumn{2}{c|}{$\spexkge_{direct}$} & \multicolumn{2}{c|}{$\spexkge_{pt}$}   & \multicolumn{2}{c|}{$\spexkge_{space}$} & \multicolumn{2}{c}{$\spexkge_{full}$} \\ \hline
			&                              & AUC            & APR            & AUC    & APR            & AUC    & APR            & AUC                  & APR        & AUC            & APR            & AUC             & APR            & AUC            & APR            \\ \hline
			\multirow{11}{*}{Valid} & 2-chain                      & 63.37          & 64.89          & 84.23  & \textbf{88.68} & 84.56  & 86.8           & 83.12                & 84.79      & \textbf{85.97} & 84.9           & 76.81           & 67.07          & 85.26          & 87.25          \\
			& 2-inter                      & 97.23          & 97.86          & 96.00  & 97.02          & 98.87  & 98.58          & 98.98                & 98.28      & 98.95          & 98.52          & 85.51           & 87.13          & \textbf{99.04} & \textbf{98.95} \\
			& Hard-2-inter                 & 70.99          & 73.55          & 66.04  & 73.83          & 73.43  & 79.98          & 73.27                & 76.36      & \textbf{74.38} & 82.16          & 63.15           & 62.91          & 73.42          & \textbf{82.52} \\
			& 3-chain                      & 61.42          & 67.94          & 79.65  & 79.45          & 79.11  & 80.93          & 77.92                & 79.26      & 79.38          & 83.97          & 70.09           & 60.8           & \textbf{80.9}  & \textbf{85.02} \\
			& 3-inter                      & 98.01          & 99.21          & 96.24  & 98.17          & 99.18  & \textbf{99.62} & \textbf{99.28}       & 99.41      & 99.1           & 99.56          & 87.62           & 89             & 99.27          & 99.59          \\
			& Hard-3-inter                 & 78.29          & 85             & 68.26  & 77.55          & 79.59  & 86.06          & 79.5                 & 84.28      & \textbf{80.48} & \textbf{87.4}  & 63.37           & 67.17          & 78.86          & 85.2           \\
			& 3-inter\_chain               & 90.56          & 94.08          & 93.39  & 91.52          & 94.59  & 90.71          & 95.99                & 95.11      & 95.86          & 94.41          & 81.16           & 83.01          & \textbf{96.7}  & \textbf{96.79} \\
			& Hard-3-inter\_chain          & 74.19          & \textbf{83.79} & 70.64  & 74.54          & 73.97  & 76.28          & 74.81                & 78.9       & \textbf{76.45} & 75.95          & 65.54           & 68.21          & 76.33          & 83.7           \\
			& 3-chain\_inter               & \textbf{98.01} & 97.45          & 92.69  & 93.31          & 96.72  & 97.61          & 97.31                & 98.67      & 97.79          & \textbf{98.76} & 83.7            & 84.42          & 97.7           & 98.65          \\
			& Hard-3-chain\_inter          & \textbf{83.59} & \textbf{88.12} & 66.86  & 74.06          & 72.12  & 77.53          & 73.23                & 79.24      & 74.74          & 80.47          & 65.13           & 69.29          & 74.72          & 78.11          \\ \hline
			& Full Valid                   & 81.57          & 85.19          & 81.4   & 84.81          & 85.21  & 87.41          & 85.34                & 87.43      & \textbf{86.31} & 88.61          & 74.21           & 73.9           & 86.22          & \textbf{89.58} \\ \hline
			\multirow{11}{*}{Test}  & 2-chain                      & 64.88          & 65.61          & 85     & 87.41          & 84.91  & 86.74          & 83.61                & 85.97      & 86.08          & 88.08          & 75.46           & 73.38          & \textbf{86.35} & \textbf{88.12} \\
			& 2-inter                      & 96.98          & 97.99          & 95.86  & 97.18          & 98.79  & 98.71          & 98.98                & 98.94      & \textbf{98.98} & \textbf{99.08} & 87.01           & 85.78          & 98.93          & 99.01          \\
			& Hard-2-inter                 & 70.39          & 76.19          & 64.5   & 71.86          & 72.15  & 79.26          & 72.04                & 79.11      & \textbf{73.72} & \textbf{81.78} & 61.22           & 62.97          & 72.62          & 81.04          \\
			& 3-chain                      & 62.3           & 62.29          & 79.19  & 80.19          & 78.93  & 80.17          & 77.53                & 78.86      & 79.43          & \textbf{81.28} & 70.55           & 68.04          & \textbf{80.49} & 80.63          \\
			& 3-inter                      & 98.09          & 99.12          & 96.54  & 97.94          & 99.33  & 99.56          & \textbf{99.45}       & 99.47      & 99.41          & \textbf{99.63} & 88.05           & 87.63          & 99.39          & 99.59          \\
			& Hard-3-inter                 & 77.27          & 83.92          & 68.69  & 75.42          & 78.93  & 83.52          & 78.58                & 84.14      & \textbf{80.11} & 84.87          & 64.44           & 64.53          & 78.76          & \textbf{84.89} \\
			& 3-inter\_chain               & 90.39          & 91.96          & 92.54  & 93.13          & 93.46  & 94.36          & 95.23                & 95.92      & 95.02          & 95.78          & 81.52           & 79.61          & \textbf{95.92} & \textbf{96.51} \\
			& Hard-3-inter\_chain          & 72.89          & 79.12          & 70.67  & 75.55          & 73.47  & 79.61          & 73.93                & 80.21      & 74.88          & 79.36          & 64.99           & 65.52          & \textbf{75.36} & \textbf{80.72} \\
			& 3-chain\_inter               & 97.35          & 98.27          & 92.22  & 94.08          & 96.55  & 96.67          & 97.29                & 98.39      & \textbf{97.79} & 98.68          & 85.28           & 84.08          & 97.64          & \textbf{98.75} \\
			& Hard-3-chain\_inter          & \textbf{83.33} & \textbf{86.24} & 66.77  & 72.1           & 72.31  & 77.89          & 73.55                & 77.08      & 75.19          & 77.42          & 65.07           & 65.41          & 74.62          & 77.31          \\ \hline
			& Full Test                    & 81.39          & 84.07          & 81.2   & 84.49          & 84.88  & 87.65          & 85.02                & 87.81      & \textbf{86.06} & 88.2           & 74.36           & 73.7           & 86.01          & \textbf{88.96} \\ 
			\bottomrule
		\end{tabular}
	}

\end{sidewaystable}

\begin{sidewaystable}[h]
	\caption{The rank of entities in the hard negative sample set of Query $\queryC$ based on $\spexkge_{full}$'s prediction for different queries: 
	1) $\bgpone$: $IsPartOf^{-1}(Alameda \; County, ?Place)$;
	2) $\bgptwo$: $Assembly(Chevrolet \; Eagle, ?Place)$;
	3) $\bgpthree$: $NearestCity(San \; Francisco \; Bay, ?Place) $;
	4) The Full Query $\queryC$.
	The correct answer is highlighted as bold.
	}
	\label{tab:qa_rank}
	\centering

\begin{tabular}{l|l|l|l|l}
\toprule
  & $\bgpone$                              & $\bgptwo$                                & $\bgpthree$                               & Query $\queryC$                                 \\ \hline
1 & dbr:Emeryville,\_California     & dbr:Flint\_Truck\_Assembly        & dbr:Alameda,\_California          & dbr:San\_Jose,\_California        \\
2 & dbr:Castro\_Valley,\_California & dbr:Norwood,\_Ohio                & dbr:San\_Jose,\_California        & \textbf{dbr:Oakland,\_California} \\
3 & dbr:Alameda,\_California        & dbr:Flint,\_Michigan              & dbr:Berkeley,\_California         & dbr:Berkeley,\_California         \\
4 & dbr:Berkeley,\_California       & dbr:Tarrytown,\_New\_York         & \textbf{dbr:Oakland,\_California} & dbr:Alameda,\_California          \\
5 & dbr:San\_Jose,\_California      & \textbf{dbr:Oakland,\_California} & dbr:Emeryville,\_California       & dbr:San\_Francisco        \\
6  & dbr:Fremont,\_California          & dbr:Alameda,\_California          & dbr:Fremont,\_California          & dbr:Fremont,\_California          \\
7  & \textbf{dbr:Oakland,\_California} & dbr:San\_Jose,\_California        & dbr:Norwood,\_Ohio                & dbr:Emeryville,\_California       \\
8  & dbr:San\_Francisco                & dbr:San\_Francisco                & dbr:San\_Francisco                & dbr:Castro\_Valley,\_California   \\
9  & dbr:Norwood,\_Ohio                & dbr:Berkeley,\_California         & dbr:Flint\_Truck\_Assembly        & dbr:Flint,\_Michigan              \\
10 & dbr:Flint\_Truck\_Assembly        & dbr:Fremont,\_California          & dbr:Flint,\_Michigan              & dbr:Norwood,\_Ohio                \\
11 & dbr:Tarrytown,\_New\_York         & dbr:Emeryville,\_California       & dbr:Castro\_Valley,\_California   & dbr:Flint\_Truck\_Assembly        \\
12  & dbr:Flint,\_Michigan              & dbr:Castro\_Valley,\_California   & dbr:Tarrytown,\_New\_York         & dbr:Tarrytown,\_New\_York \\

\bottomrule
\end{tabular}
\end{sidewaystable}

\begin{figure*}[t!]
	\centering \tiny
	\begin{subfigure}[b]{0.49\textwidth}  
		\centering 
		\includegraphics[width=\textwidth]{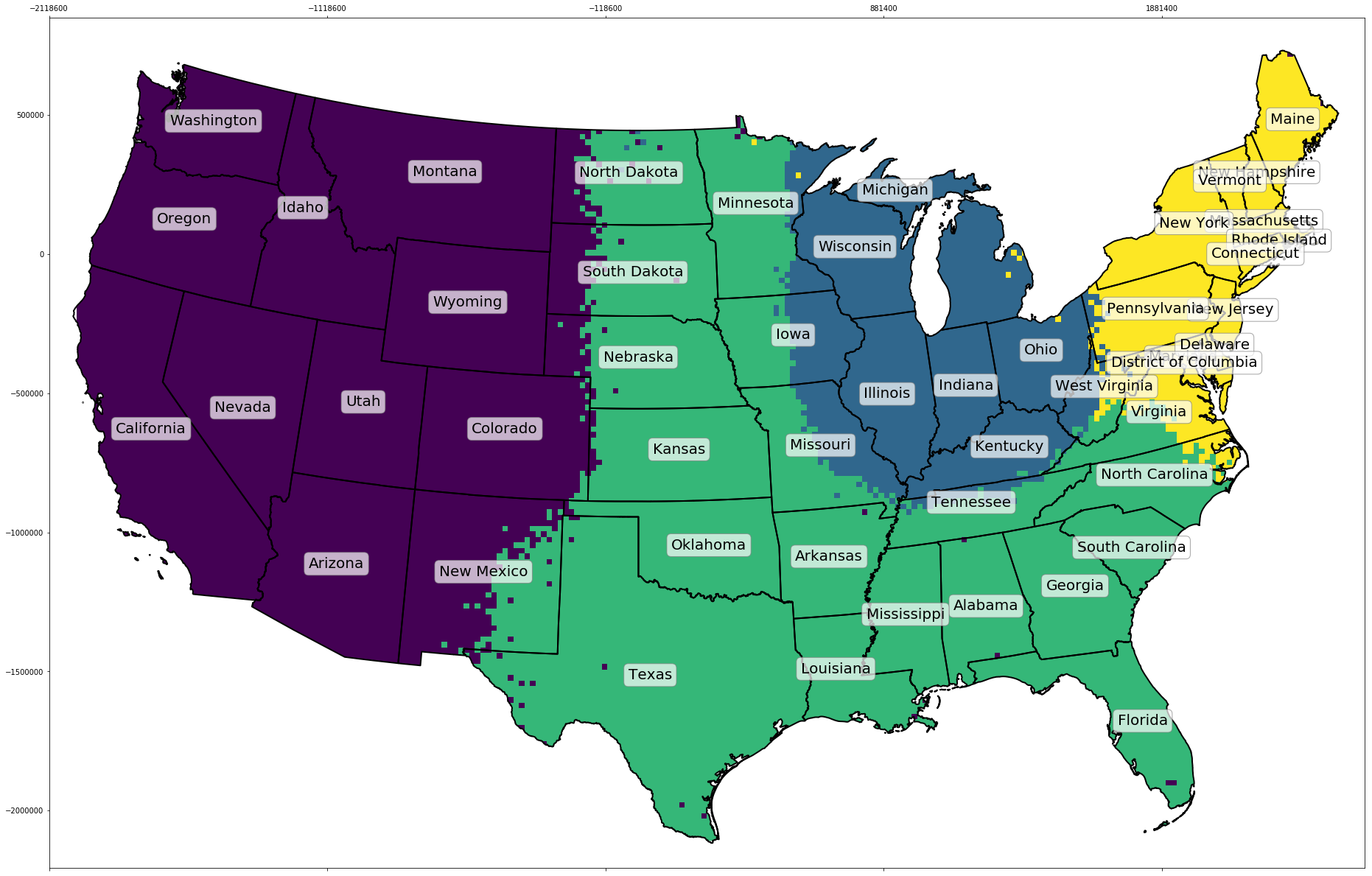}\vspace*{-0.1cm}
		\caption[]
		{{ 
		}}    
		\label{fig:res_map_clustering}
	\end{subfigure}
	\hfill
	\begin{subfigure}[b]{0.49\textwidth}  
		\centering 
		\includegraphics[width=\textwidth]{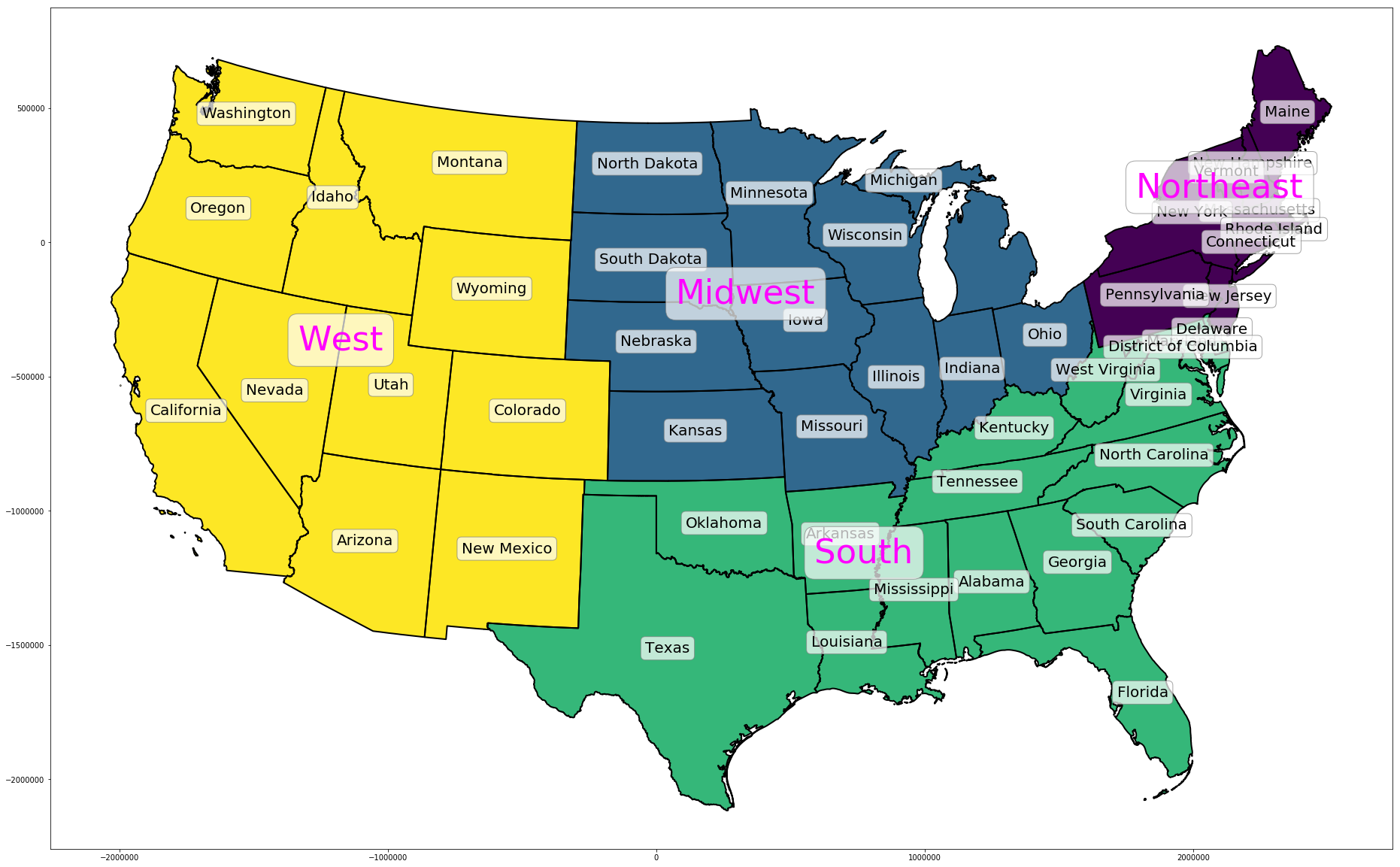}\vspace*{-0.1cm}
		\caption[]
		{{
		}}    
		\label{fig:usregions}
	\end{subfigure}
	\begin{subfigure}[b]{0.49\textwidth}  
		\centering 
		\includegraphics[width=\textwidth]{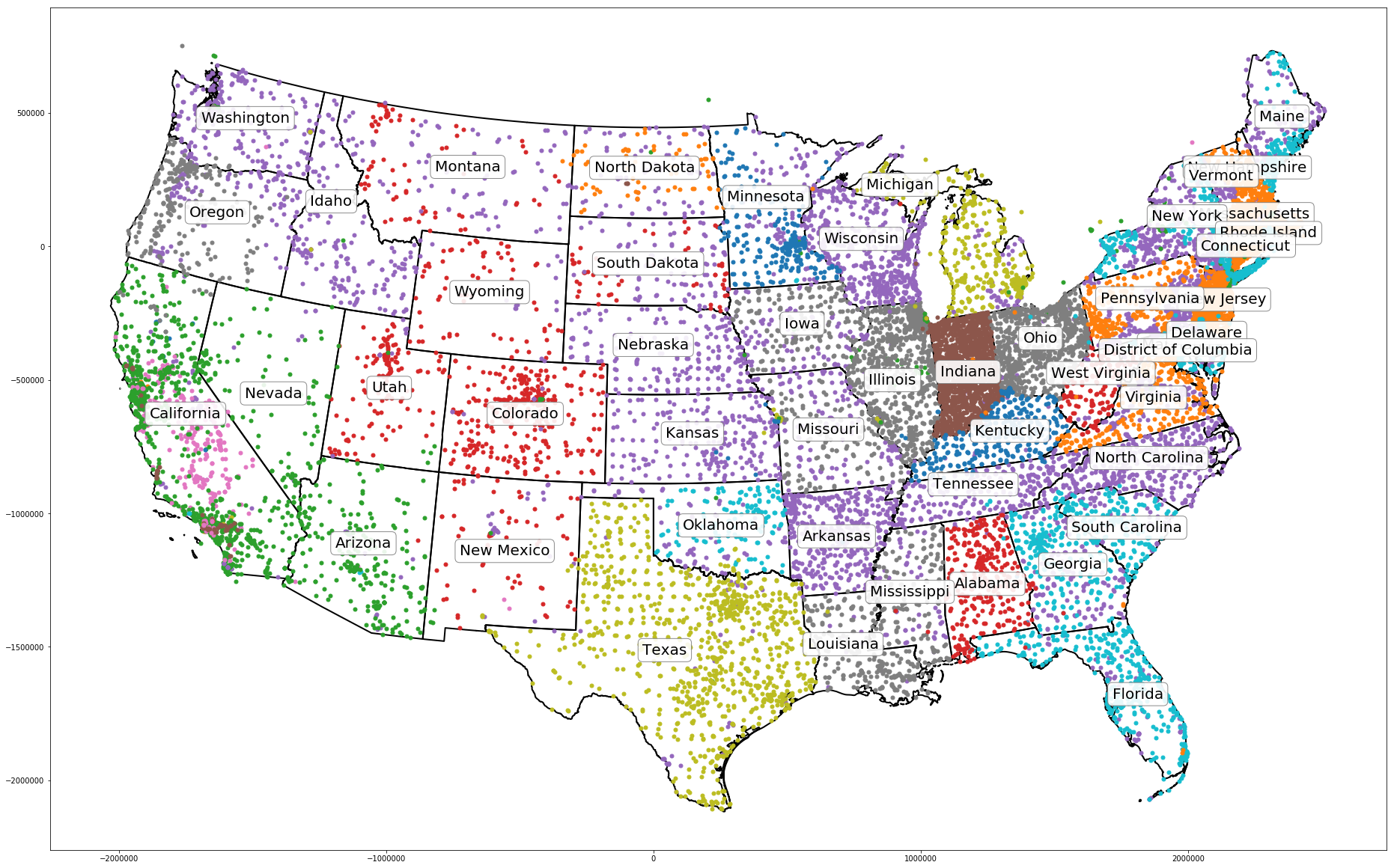}\vspace*{-0.1cm}
		\caption[]
		{{ 
		}}    
		\label{fig:kgclustering}
	\end{subfigure}
	
	\caption{
	\small
	 (a) Clustering result of location embeddings produced by the location encoder $\locenc()$ in 
	$\spexkge_{space}$. It illustrates spatial coherence and semantics (b) Census Bureau-designated regions of United States, and (c) the community detection (Shuffled Louvain) results of knowledge graph $\kg$ by treating $\kg$ as a undirected unlabeled multigraph. It lacks spatial coherence.
	} 
	\label{fig:res_map}
	\vspace*{-0.15cm}
\end{figure*}

We also test how well the location encoder $\locenc()$ in $\spexkge$ can capture the global position information and how $\locenc()$ interacts with other components of $\spexkge$. We use $\spexkge_{space}$ as an example.
Since $\locenc()$ is an inductive learning model, we divide the study area $\studyarea$ into $20km \times 20km$ grids and take the location of each grid center as the input of $\locenc()$. Each grid will get a $\pedim$ dimension location embedding after location encoding. We apply hierarchical clustering on these embeddings. Figure \ref{fig:res_map_clustering} shows the clustering result. We compare it with the widely used USA Census Bureau-designated regions\footnote{\url{https://en.wikipedia.org/wiki/List_of_regions_of_the_United_States}} (See Figure \ref{fig:usregions}). We can see that Figure \ref{fig:res_map_clustering} and \ref{fig:usregions} look very similar to each other. We use two clustering evaluation metrics - Normalized Mutual Information (NMI) and Rand Index - to measure the degree of similarity which yield 0.62 on NMI and 0.63 on Rand Index. To take a closer look at Figure \ref{fig:res_map_clustering}, we can also see that the clusters are divided on the state borders. We hypothesize that this is because $\locenc()$ is informed of the connectivity of different geographic entities in $\kg$ during model training, resulting in that locations which are connected in original $\kg$ are also clustered after training.

To validate this hypothesis, we apply Louvain community detection algorithm with a shuffled node sequence\footnote{\url{https://github.com/tsakim/Shuffled_Louvain}} on the original $\kg$ by treating $\kg$ as an undirected and unlabeled graph. Figure \ref{fig:kgclustering} shows the community structure with the best modularity which contains 32 communities. Some interesting observations can be made by comparing these three figures: 
\begin{enumerate}
    \item Most communities in Figure \ref{fig:kgclustering} are separated at state borders, which is an evidence of our hypothesis;
    \item Some communities contain locations at different states, which are far away from each other. For example, the red community which contains locations from Utah, Colorado, and Alabama. This indicates that some locations are very similar purely based on the graph structure of $\kg$ . As $\locenc()$ imposes spatial constraints on entities, spatially coherent clusters in Figure \ref{fig:res_map_clustering} are presented.
\end{enumerate}

One hypothesis why Figure \ref{fig:res_map_clustering} and \ref{fig:usregions} look similar is that in the KG, the number of connections between entities within one Bureau-designated region is more than the number of connections among entities in different regions. This may be due to the fact that DBpedia uses census data as one of the data sources while census data is organized in a way which reflects Bureau-designated regions of the US. More research is needed to validate this hypothesis in the future.

 \subsection{Evaluation on Spatial Semantic Lifting Task}  \label{subsec:spasemlift_eval}

\subsubsection{Baselines}  \label{subsubsec:ssl_baseline}
The spatial semantic lifting model is composed of $\enc()$, $\projent()$, and $\projloc()$ which is indicated as $\spexkge_{\ssl}$. 
In order to 
study the contribution of feature encoder  and location encoder,
we create a baseline $\spexkge_{space}^{\prime}$ whose entity encoder does not have the feature encoder component, similar to $\spexkge_{space}$. The difference is that they are trained on different objectives. These are the only two models that can do spatial semantic lifting task, since they are fully inductive learning models directly using locations as the only input features.

\subsubsection{Training Detail}  \label{subsubsec:ssl_train_detail}
We train $\spexkge_{\ssl}$ and $\spexkge_{space}^{\prime}$ based on $\loss^{(SSL)}$. To quantitatively evaluate them on spatial semantic lifting task, we use $\triset_{s} \cap \triset_{o}$ in the validation and testing dataset with different relations (See Table \ref{tab:data}). For each triple $\triple_{i} = (h_{i},\rel_{i},t_{i}) \in \triset_{s}$, given the head entity's location and $\rel_{i}$, we use $\projloc(\geofun(h_{i}),\rel_{i})$ (See Equation~\ref{equ:ent2x}) to predict the tail entity embedding. Similar process can be done for $\triple_{j} = (h_{j},\rel_{j},t_{j}) \in \triset_{o}$ but from the reverse direction.  We also use AUC and APR as the evaluation metrics. Note that since $\geofun(h_{i}) = \bx_{i}^{(t)} \sim \uniondist(\bx_{i}^{min},\bx_{i}^{max}) 
\; ,  \;  \pgonfun(h_{i}) = [\bx_{i}^{min};\bx_{i}^{max}]  \; if \; h_{i} \in \pgonentset$, the location of head entity is randomly generated, which can be treated as unseen in the training process. We use the same hyperparameter configuration as $\spexkge_{full}$.

\begin{table}[t!]
	\caption{The evaluation of spatial semantic lifting on $\dbgeo$ over all validation/testing triples  
	}
	\label{tab:ssl_eval}
	\centering
\begin{tabular}{l|cc|cc|cc}
\toprule
      & \multicolumn{2}{c|}{$\spexkge_{space}$} & \multicolumn{2}{c|}{$\spexkge_{\ssl}$}   & \multicolumn{2}{c}{$\spexkge_{\ssl}$ - $\spexkge_{space}$} \\ \hline
      & AUC            & APR            & AUC            & APR            & $\Delta$AUC         & $\Delta$APR         \\ \hline
Valid & 72.85          & 75.49          & \textbf{82.74} & \textbf{85.51} & 9.89        & 10.02       \\
Test  & 73.41          & 75.77          & \textbf{83.27} & \textbf{85.36} & 9.86        & 9.59       \\
\bottomrule
\end{tabular}
\end{table}

\subsubsection{Evaluation Results}  \label{subsubsec:ssl_eval_res}
Table \ref{tab:ssl_eval} shows the overall evaluation results. We can see that $\spexkge_{\ssl}$ outperforms $\spexkge_{space}^{\prime}$ with a significant margin ($\Delta$AUC = 9.86\% and $\Delta$APR = 9.59\% on the testing dataset) which clearly shows the strength of considering both feature embedding and space embedding in spatial semantic lifting task.

Next, among all validation and testing triples with different relations, we select a few relations and report APR of two models on these triples with specific relations. The results are shown in Table \ref{tab:ssl_evalperq}. These relations are selected since they are interesting from spatial reasoning perspective. We can see that $\spexkge_{\ssl}$ outperforms $\spexkge_{space}^{\prime}$ on all these triple sets with different relations.

\begin{table}[t!]
	\caption{The evaluation of $\spexkge_{\ssl}$ and $\spexkge_{space}^{\prime}$ on $\dbgeo$ for a few selected relation $\rel$ (using APR (\%) as evaluation metric).}
	\label{tab:ssl_evalperq}
	\centering
\begin{tabular}{l|l|c|c|c}
\toprule
                       & Query Type     & $\spexkge_{space}^{\prime}$ & $\spexkge_{\ssl}$      & $\Delta$APR \\ \hline
\multirow{7}{*}{Valid} & $state(\bx,?e)$                  & 92.00       & \textbf{99.94} & 7.94  \\
                       & $nearestCity(\bx,?e)$            & 84.00       & \textbf{94.00} & 10.00    \\
                       & $broadcastArea^{-1}(\bx,?e)$     & 91.60       & \textbf{95.60} & 4.00     \\
                       & $isPartOf(\bx,?e)$               & 88.56       & \textbf{98.88} & 10.32 \\
                       & $locationCity(\bx,?e)$           & 83.50       & \textbf{99.00} & 15.50  \\
                       & $residence^{-1}(\bx,?e)$         & 90.50       & \textbf{93.50} & 3.00     \\
                       & $hometown^{-1}(\bx,?e)$          & 61.14       & \textbf{74.86} & 13.71 \\ \hline
\multirow{7}{*}{Test}  & $state(\bx,?e)$                  & 89.06       & \textbf{99.97} & 10.91 \\
                       & $nearestCity(\bx,?e)$            & 87.60       & \textbf{99.80} & 12.20  \\
                       & $broadcastArea^{-1}(\bx,?e)$     & 90.81       & \textbf{96.63} & 5.82  \\
                       & $isPartOf(\bx,?e)$               & 87.66       & \textbf{98.87} & 11.21 \\
                       & $locationCity(\bx,?e)$           & 84.80       & \textbf{99.10} & 14.30  \\
                       & $residence^{-1}(\bx,?e)$         & 61.21       & \textbf{77.68} & 16.47 \\
                       & $hometown^{-1}(\bx,?e)$          & 61.44       & \textbf{76.83} & 15.39 \\
                       \bottomrule
\end{tabular}
\end{table}

\begin{figure*}[t!]
	\centering \tiny
    \begin{subfigure}[b]{0.32\textwidth}  
		\centering 
		\includegraphics[width=\textwidth]{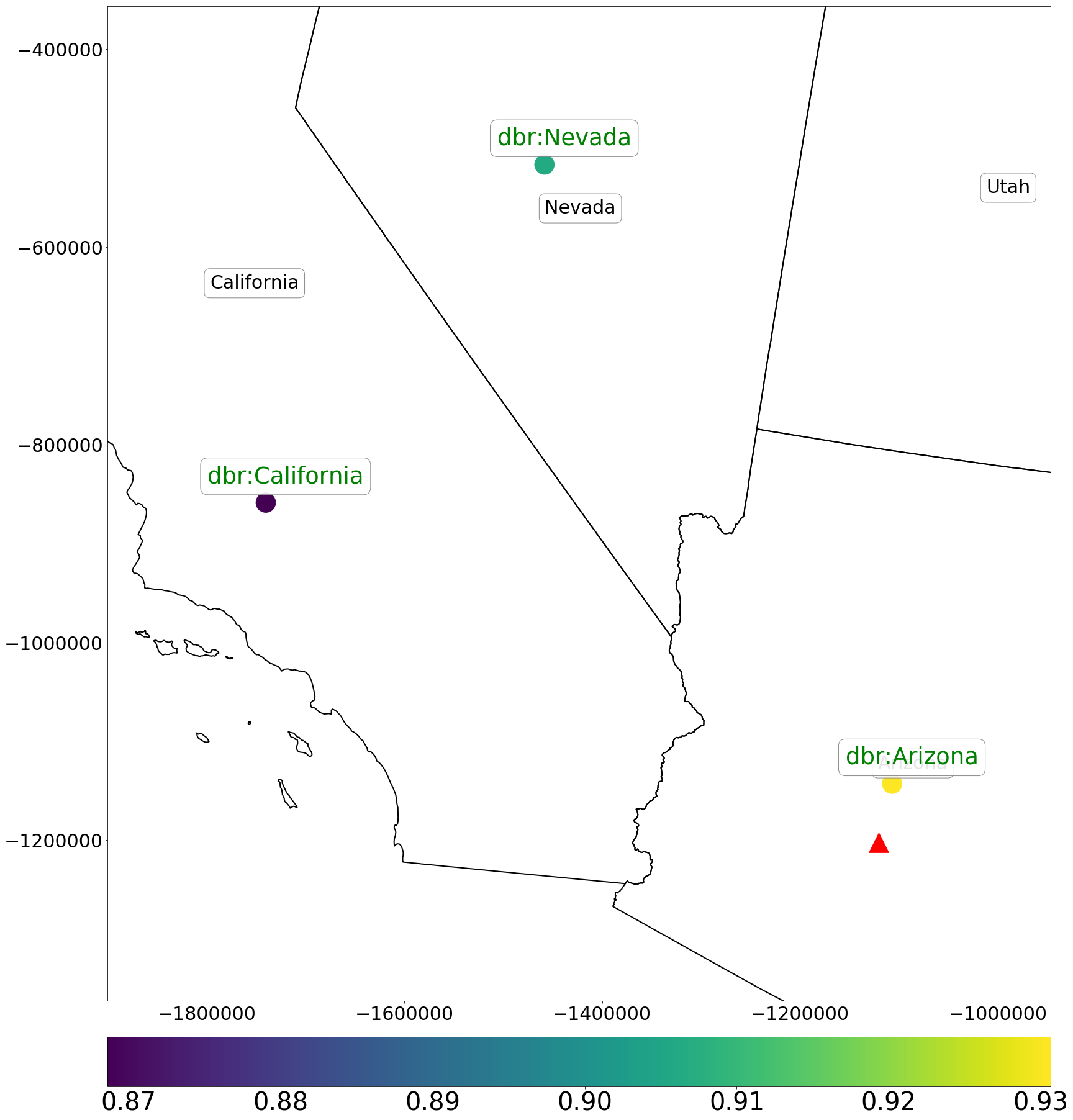}\vspace*{-0.1cm}
		\caption[]
		{{ 
		$state(\bx,?e)$
		}}    
		\label{fig:state}
	\end{subfigure}
	\begin{subfigure}[b]{0.325\textwidth}  
		\centering 
		\includegraphics[width=\textwidth]{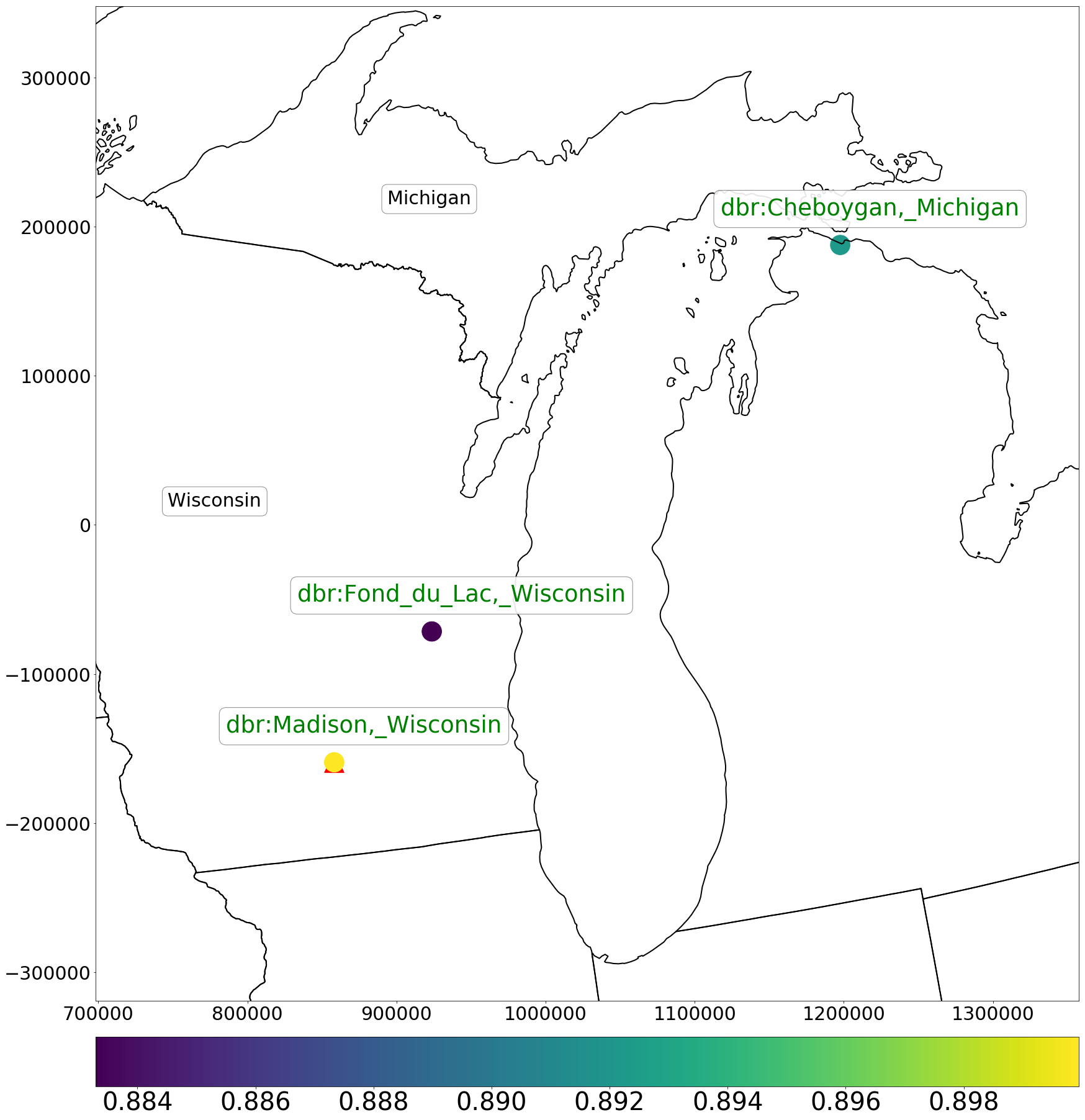}\vspace*{-0.1cm}
		\caption[]
		{{
		$nearestCity(\bx,?e)$
		}}    
		\label{fig:nearestcity}
	\end{subfigure}
	\begin{subfigure}[b]{0.325\textwidth}  
		\centering 
		\includegraphics[width=\textwidth]{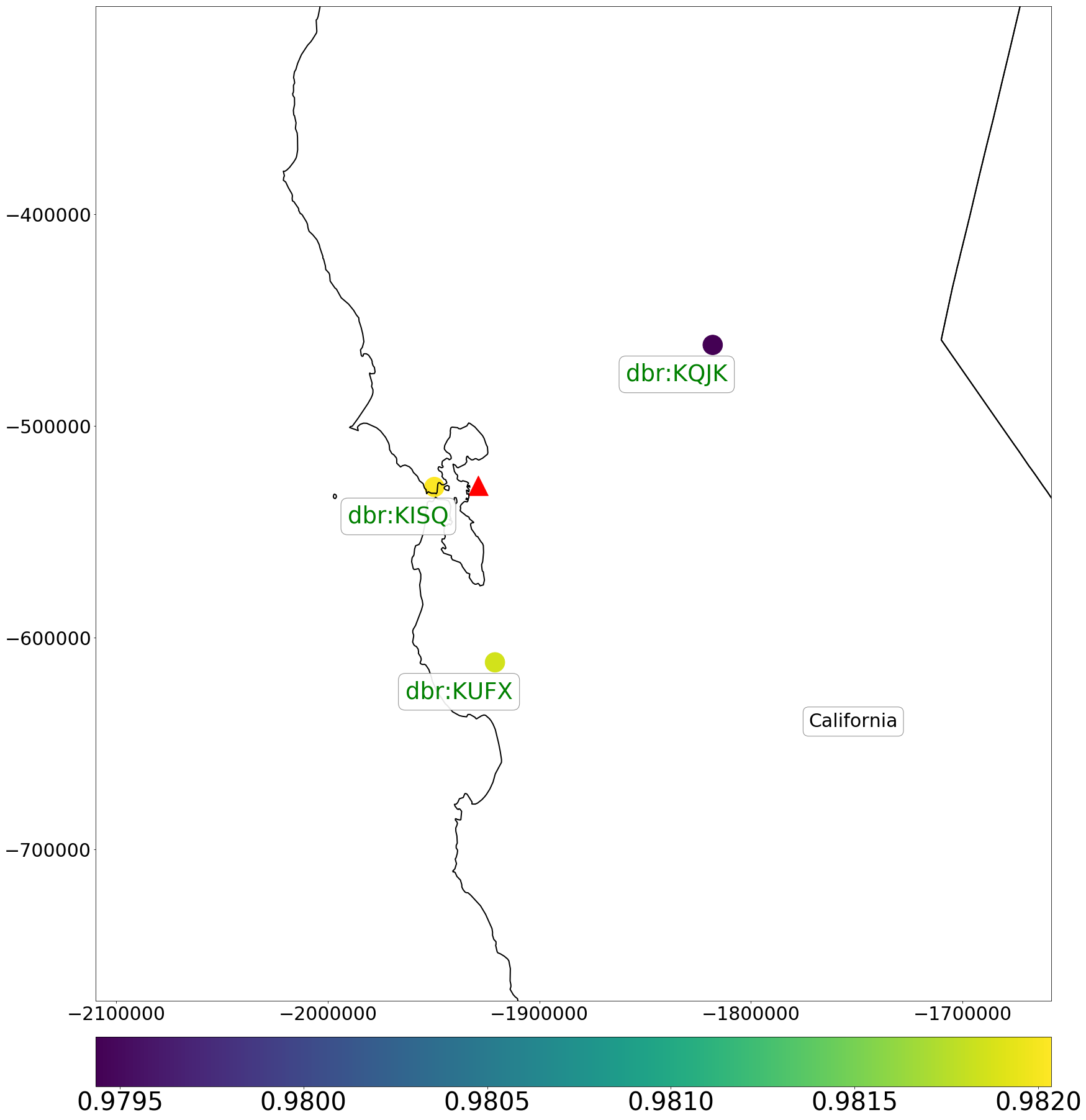}\vspace*{-0.1cm}
		\caption[]
		{{ 
	    $broadcastArea^{-1}(\bx,?e)$
		}}    
		\label{fig:broadcastarea}
	\end{subfigure}
	\begin{subfigure}[b]{0.32\textwidth}  
		\centering 
		\includegraphics[width=\textwidth]{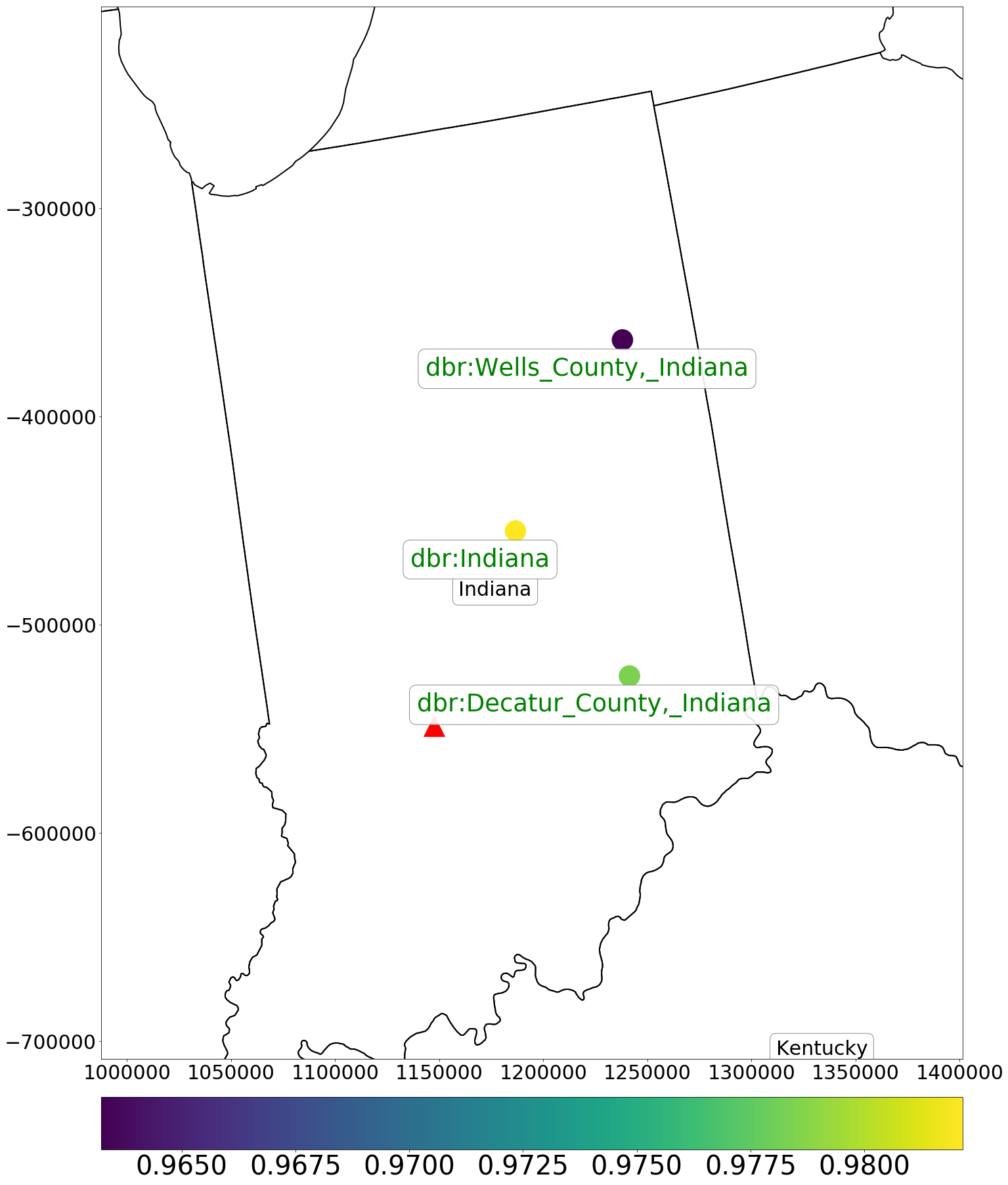}\vspace*{-0.1cm}
		\caption[]
		{{ 
		$isPartOf(\bx,?e)$
		}}    
		\label{fig:ispartof}
	\end{subfigure}
	\caption{
	\small
	The visualization of spatial semantic lifting of $\spexkge_{\ssl}$. Figure (a), (b), (c), and (d) shows the top 3 geographic entities which can answer query $\rel(\bx,?e)$ where $\rel$ is the relation we pick. 
	\textbf{Red triangle}: the select location $\bx$.
	\textbf{Circles}: top 3 geographic entities ranked by our model, and their
	colors indicates cosine similarity between the geographic entities and the predicted query embedding.
	} 
	\label{fig:spa_sem_lift_eval}
	\vspace*{-0.15cm}
\end{figure*}
In order to know how well $\spexkge_{\ssl}$ understands the semantics of different types of (spatial) relations, we visualize the spatial semantic lifting results in Figure \ref{fig:spa_sem_lift_eval} for four spatial relations: \texttt{dbo:state}, \texttt{dbo:nearestCity}, , $\texttt{dbo:broadcastArea}^{-1}$, and \texttt{dbo:isPartOf}.
\texttt{dbo:state}, \texttt{dbo:isPartOf}, and $\texttt{dbo:broadcastArea}^{-1}$ are about partonomy relations while \texttt{dbo:nearestCity} represents an example of point-wise metric spatial relations. 
Some interesting observations can be made:
\begin{enumerate}
    \item $\spexkge_{\ssl}$ is capable of capturing the spatial proximity such that the top 1 geographic entity (yellow point) in each case is the closest to location $\bx$ (red triangle). We also treat this as an indicator for the capability of $\spexkge_{\ssl}$ to handle partonomy relations and point-wise metric spatial relations.
    \item $\spexkge_{\ssl}$ can capture the semantics of relations, e.g., the domain and range of each relation/predicate. All top ranked entities are within the range of the corresponding relation.
    For example, in Figure \ref{fig:state} with query $state(\bx,?e)$, the top 3 entities are all states spatially close to $\bx$. In Figure \ref{fig:broadcastarea} with query $broadcastArea^{-1}(\bx,?e)$, all top 3 entities are nearby radio stations. In Figure \ref{fig:ispartof} (d) with query $isPartOf(\bx,?e)$, all top 3 entities are states (\texttt{dbo:Indiana}) and counties.
    \item We notice that the result of query $nearestCity(\bx,?e)$ in Figure \ref{fig:nearestcity} is not good enough since the second result - \texttt{dbo:Cheboygan,\_Michigan} - is outside of Wisconsin. After investigating the triples with \texttt{dbo:nearestCity} as the relation, we find out \texttt{dbo:nearestCity} usually links a natural resource entity (e.g., lakes, national parks) to a city. These natural resource entities usually cover large area and complex geometries. So \texttt{dbo:nearestCity} is not a purely point-wise distance base relation but a complex distance base relation based on their real geometries. Since our model only takes the bounding box of each entity and there are usually no subdivions of these nature resource entities, it is hard for our model to learn the semantics of \texttt{dbo:nearestCity}.

\end{enumerate}

Based on the evaluation results and model analysis, we can see that given a relation $\rel$, $\spexkge_{\ssl}$ is able to link a location $\bx$ to an entity $\ent$ in $\kg$ by considering the semantics of $\rel$ and spatial proximity.

  \section{Conclusion} \label{sec:conclusion}
In this work, we propose a location-aware knowledge graph embedding model called $\spexkge$ which  
enables spatial reasoning in the embedding space for its 
three major components - entity embedding encoder $\enc()$, projection operator $\proj()$, and intersection operator $\inter()$.
We demonstrate how to incorporate spatial information of geographic entities such as locations and spatial extents into $\enc()$ such that $\spexkge$ can handle different types of spatial relations such as point-wise metric spatial relations and partonomy relations. To the best of our knowledge, this is the first KG embedding model which incorporates location encoding into the model architecture instead of relying on some form of distance measure among entities while capturing the scale effect of different geographic entities. 
Two tasks have been used to evaluate the performance of $\spexkge$ - geographic logic query answering and spatial semantic lifting. Results show that $\spexkge_{full}$ can outperform multiple baselines on the geographic logic query answering task which indicates the effectiveness of spatially explicit models. It also demonstrates the importance to considering the scale effect in location encoding. 
Also we proposed a new task - spatial semantic lifting, aiming at linking a randomly selected location to entities in the existing KG with some relation. None of the existing KG embedding models can solve this task except our model. 
We have shown that  $\spexkge_{\ssl}$ can significantly outperform the baseline $\spexkge_{space}^{\prime}$  ($\Delta$AUC = 9.86\% and $\Delta$APR = 9.59\% on the testing dataset). Visualizations show that $\spexkge_{\ssl}$ can successfully capture the spatial proximity information as well as the semantics of relations. 
In the future, we hope to explore a more concise way to encode the spatial footprints of geographic entities in a KG. Moreover, we want to explore more varieties of the spatial semantic lifting task. %

\addcontentsline{toc}{section}{References}

\bibliographystyle{havard}


\begin{thebibliography}{43}
	\expandafter\ifx\csname natexlab\endcsname\relax\def\natexlab#1{#1}\fi
	\expandafter\ifx\csname url\endcsname\relax
	\def\url#1{\texttt{#1}}\fi
	\expandafter\ifx\csname urlprefix\endcsname\relax\def\urlprefix{URL }\fi
	
	\bibitem[{Abbott and Callaway(2014)}]{abbott2014nobel}
	Abbott, A., Callaway, E., 2014. Nobel prize for decoding brain's sense of
	place. Nature 514~(7521).
	
	\bibitem[{Battaglia et~al.(2018)Battaglia, Hamrick, Bapst, Sanchez-Gonzalez,
		Zambaldi, Malinowski, Tacchetti, Raposo, Santoro, Faulkner,
		et~al.}]{battaglia2018relational}
	Battaglia, P.~W., Hamrick, J.~B., Bapst, V., Sanchez-Gonzalez, A., Zambaldi,
	V., Malinowski, M., Tacchetti, A., Raposo, D., Santoro, A., Faulkner, R.,
	et~al., 2018. Relational inductive biases, deep learning, and graph networks.
	arXiv preprint arXiv:1806.01261 .
	
	\bibitem[{Berant et~al.(2013)Berant, Chou, Frostig, and
		Liang}]{berant2013semantic}
	Berant, J., Chou, A., Frostig, R., Liang, P., 2013. Semantic parsing on
	freebase from question-answer pairs. In: Proceedings of the 2013 conference
	on empirical methods in natural language processing. pp. 1533--1544.
	
	\bibitem[{Bordes et~al.(2013)Bordes, Usunier, Garcia-Duran, Weston, and
		Yakhnenko}]{bordes2013translating}
	Bordes, A., Usunier, N., Garcia-Duran, A., Weston, J., Yakhnenko, O., 2013.
	Translating embeddings for modeling multi-relational data. In: Advances in
	neural information processing systems. pp. 2787--2795.
	
	\bibitem[{Cai et~al.(2019)Cai, Yan, Mai, Janowicz, and Zhu}]{cai2019transgcn}
	Cai, L., Yan, B., Mai, G., Janowicz, K., Zhu, R., 2019. Transgcn: Coupling
	transformation assumptions with graph convolutional networks for link
	prediction. In: K-CAP. ACM.
	
	\bibitem[{Chu et~al.(2019)Chu, Potetz, Wang, Howard, Song, Brucher, Leung, and
		Adam}]{chu2019geo}
	Chu, G., Potetz, B., Wang, W., Howard, A., Song, Y., Brucher, F., Leung, T.,
	Adam, H., 2019. Geo-aware networks for fine-grained recognition. In:
	Proceedings of the IEEE International Conference on Computer Vision
	Workshops. pp. 0--0.
	
	\bibitem[{De~Nicola et~al.(2008)De~Nicola, Di~Mascio, Lezoche, and
		Tagliano}]{de2008semantic}
	De~Nicola, A., Di~Mascio, T., Lezoche, M., Tagliano, F., 2008. Semantic lifting
	of business process models. In: 2008 12th Enterprise Distributed Object
	Computing Conference Workshops. IEEE, pp. 120--126.
	
	\bibitem[{Dong et~al.(2014)Dong, Gabrilovich, Heitz, Horn, Lao, Murphy,
		Strohmann, Sun, and Zhang}]{dong2014knowledge}
	Dong, X., Gabrilovich, E., Heitz, G., Horn, W., Lao, N., Murphy, K., Strohmann,
	T., Sun, S., Zhang, W., 2014. Knowledge vault: A web-scale approach to
	probabilistic knowledge fusion. In: Proceedings of the 20th ACM SIGKDD
	international conference on Knowledge discovery and data mining. ACM, pp.
	601--610.
	
	\bibitem[{Hamilton et~al.(2018)Hamilton, Bajaj, Zitnik, Jurafsky, and
		Leskovec}]{hamilton2018embedding}
	Hamilton, W., Bajaj, P., Zitnik, M., Jurafsky, D., Leskovec, J., 2018.
	Embedding logical queries on knowledge graphs. In: Advances in Neural
	Information Processing Systems. pp. 2026--2037.
	
	\bibitem[{Hamilton et~al.(2017{\natexlab{a}})Hamilton, Ying, and
		Leskovec}]{hamilton2017inductive}
	Hamilton, W., Ying, Z., Leskovec, J., 2017{\natexlab{a}}. Inductive
	representation learning on large graphs. In: Advances in Neural Information
	Processing Systems. pp. 1024--1034.
	
	\bibitem[{Hamilton et~al.(2017{\natexlab{b}})Hamilton, Ying, and
		Leskovec}]{hamilton2017representation}
	Hamilton, W.~L., Ying, R., Leskovec, J., 2017{\natexlab{b}}. Representation
	learning on graphs: Methods and applications. arXiv preprint arXiv:1709.05584
	.
	
	\bibitem[{Hoffart et~al.(2013)Hoffart, Suchanek, Berberich, and
		Weikum}]{hoffart2013yago2}
	Hoffart, J., Suchanek, F.~M., Berberich, K., Weikum, G., 2013. Yago2: A
	spatially and temporally enhanced knowledge base from wikipedia. Artificial
	Intelligence 194, 28--61.
	
	\bibitem[{Janowicz et~al.(2016)Janowicz, Hu, McKenzie, Gao, Regalia, Mai, Zhu,
		Adams, and Taylor}]{janowicz2016moon}
	Janowicz, K., Hu, Y., McKenzie, G., Gao, S., Regalia, B., Mai, G., Zhu, R.,
	Adams, B., Taylor, K., 2016. Moon landing or safari? a study of systematic
	errors and their causes in geographic linked data. In: The Annual
	International Conference on Geographic Information Science. Springer, pp.
	275--290.
	
	\bibitem[{Janowicz et~al.(2012)Janowicz, Scheider, Pehle, and
		Hart}]{janowicz2012geospatial}
	Janowicz, K., Scheider, S., Pehle, T., Hart, G., 2012. Geospatial semantics and
	linked spatiotemporal data--past, present, and future. Semantic Web 3~(4),
	321--332.
	
	\bibitem[{Kejriwal and Szekely(2017)}]{kejriwal2017neural}
	Kejriwal, M., Szekely, P., 2017. Neural embeddings for populated geonames
	locations. In: International Semantic Web Conference. Springer, pp. 139--146.
	
	\bibitem[{Kristiadi et~al.(2019)Kristiadi, Khan, Lukovnikov, Lehmann, and
		Fischer}]{kristiadi2019incorporating}
	Kristiadi, A., Khan, M.~A., Lukovnikov, D., Lehmann, J., Fischer, A., 2019.
	Incorporating literals into knowledge graph embeddings. In: International
	Semantic Web Conference. Springer, pp. 347--363.
	
	\bibitem[{Lao et~al.(2011)Lao, Mitchell, and Cohen}]{lao2011random}
	Lao, N., Mitchell, T., Cohen, W.~W., Jul. 2011. Random walk inference and
	learning in a large scale knowledge base. In: Proceedings of the 2011
	Conference on Empirical Methods in Natural Language Processing. Association
	for Computational Linguistics, Edinburgh, Scotland, UK., pp. 529--539.
	
	\bibitem[{Liang et~al.(2017)Liang, Berant, Le, Forbus, and
		Lao}]{liang2017neural}
	Liang, C., Berant, J., Le, Q., Forbus, K., Lao, N., 2017. Neural symbolic
	machines: Learning semantic parsers on freebase with weak supervision. In:
	Proceedings of the 55th Annual Meeting of the Association for Computational
	Linguistics (Volume 1: Long Papers). pp. 23--33.
	
	\bibitem[{Mac~Aodha et~al.(2019)Mac~Aodha, Cole, and Perona}]{mac2019presence}
	Mac~Aodha, O., Cole, E., Perona, P., 2019. Presence-only geographical priors
	for fine-grained image classification. In: Proceedings of the IEEE
	International Conference on Computer Vision. pp. 9596--9606.
	
	\bibitem[{Mai et~al.(2018)Mai, Janowicz, and Yan}]{mai2018support}
	Mai, G., Janowicz, K., Yan, B., 2018. Support and centrality: Learning weights
	for knowledge graph embedding models. In: EKAW. Springer, pp. 212--227.
	
	\bibitem[{Mai et~al.(2019{\natexlab{a}})Mai, Janowicz, Yan, Zhu, Cai, and
		Lao}]{mai2019contextual}
	Mai, G., Janowicz, K., Yan, B., Zhu, R., Cai, L., Lao, N., 2019{\natexlab{a}}.
	Contextual graph attention for answering logical queries over incomplete
	knowledge graphs. In: K-CAP. ACM.
	
	\bibitem[{Mai et~al.(2020)Mai, Janowicz, Yan, Zhu, Cai, and
		Lao}]{mai2020multiscale}
	Mai, G., Janowicz, K., Yan, B., Zhu, R., Cai, L., Lao, N., 2020. Multi-scale
	representation learning for spatial feature distributions using grid cells.
	In: The Eighth International Conference on Learning Representations.
	openreview.
	
	\bibitem[{Mai et~al.(2019{\natexlab{b}})Mai, Yan, Janowicz, and
		Zhu}]{mai2019relaxing}
	Mai, G., Yan, B., Janowicz, K., Zhu, R., 2019{\natexlab{b}}. Relaxing
	unanswerable geographic questions using a spatially explicit knowledge graph
	embedding model. In: AGILE 2019. Springer, pp. 21--39.
	
	\bibitem[{Nickel et~al.(2015)Nickel, Murphy, Tresp, and
		Gabrilovich}]{nickel2015review}
	Nickel, M., Murphy, K., Tresp, V., Gabrilovich, E., 2015. A review of
	relational machine learning for knowledge graphs. Proceedings of the IEEE
	104~(1), 11--33.
	
	\bibitem[{Nickel et~al.(2016)Nickel, Rosasco, Poggio,
		et~al.}]{nickel2016holographic}
	Nickel, M., Rosasco, L., Poggio, T.~A., et~al., 2016. Holographic embeddings of
	knowledge graphs. In: AAAI. pp. 1955--1961.
	
	\bibitem[{Nickel et~al.(2012)Nickel, Tresp, and
		Kriegel}]{nickel2012factorizing}
	Nickel, M., Tresp, V., Kriegel, H.-P., 2012. Factorizing yago: scalable machine
	learning for linked data. In: WWW. ACM, pp. 271--280.
	
	\bibitem[{Pennington et~al.(2014)Pennington, Socher, and
		Manning}]{pennington2014glove}
	Pennington, J., Socher, R., Manning, C.~D., 2014. Glove: Global vectors for
	word representation. In: Proceedings of the 2014 conference on empirical
	methods in natural language processing (EMNLP). pp. 1532--1543.
	
	\bibitem[{Pezeshkpour et~al.(2018)Pezeshkpour, Chen, and
		Singh}]{pezeshkpour2018embedding}
	Pezeshkpour, P., Chen, L., Singh, S., 2018. Embedding multimodal relational
	data for knowledge base completion. In: Proceedings of the 2018 Conference on
	Empirical Methods in Natural Language Processing. pp. 3208--3218.
	
	\bibitem[{Qiu et~al.(2019)Qiu, Gao, Yu, and Lu}]{qiu2019knowledge}
	Qiu, P., Gao, J., Yu, L., Lu, F., 2019. Knowledge embedding with geospatial
	distance restriction for geographic knowledge graph completion. ISPRS
	International Journal of Geo-Information 8~(6), 254.
	
	\bibitem[{Rajpurkar et~al.(2018)Rajpurkar, Jia, and Liang}]{rajpurkar2018know}
	Rajpurkar, P., Jia, R., Liang, P., 2018. Know what you don’t know:
	Unanswerable questions for squad. In: Proceedings of the 56th Annual Meeting
	of the Association for Computational Linguistics (Volume 2: Short Papers).
	pp. 784--789.
	
	\bibitem[{Regalia et~al.(2019)Regalia, Janowicz, and
		McKenzie}]{regalia2019computing}
	Regalia, B., Janowicz, K., McKenzie, G., 2019. Computing and querying strict,
	approximate, and metrically refined topological relations in linked
	geographic data. Transactions in GIS 23~(3), 601--619.
	
	\bibitem[{Schlichtkrull et~al.(2018)Schlichtkrull, Kipf, Bloem, Van Den~Berg,
		Titov, and Welling}]{schlichtkrull2018modeling}
	Schlichtkrull, M., Kipf, T.~N., Bloem, P., Van Den~Berg, R., Titov, I.,
	Welling, M., 2018. Modeling relational data with graph convolutional
	networks. In: European Semantic Web Conference. Springer, pp. 593--607.
	
	\bibitem[{Sch{\"o}lkopf(2001)}]{scholkopf2001kernel}
	Sch{\"o}lkopf, B., 2001. The kernel trick for distances. In: Advances in neural
	information processing systems. pp. 301--307.
	
	\bibitem[{Scholkopf et~al.(1997)Scholkopf, Sung, Burges, Girosi, Niyogi,
		Poggio, and Vapnik}]{scholkopf1997comparing}
	Scholkopf, B., Sung, K.-K., Burges, C.~J., Girosi, F., Niyogi, P., Poggio, T.,
	Vapnik, V., 1997. Comparing support vector machines with gaussian kernels to
	radial basis function classifiers. IEEE transactions on Signal Processing
	45~(11), 2758--2765.
	
	\bibitem[{Trisedya et~al.(2019)Trisedya, Qi, and Zhang}]{trisedya2019entity}
	Trisedya, B.~D., Qi, J., Zhang, R., 2019. Entity alignment between knowledge
	graphs using attribute embeddings. In: Proceedings of the AAAI Conference on
	Artificial Intelligence. Vol.~33. pp. 297--304.
	
	\bibitem[{Vaswani et~al.(2017)Vaswani, Shazeer, Parmar, Uszkoreit, Jones,
		Gomez, Kaiser, and Polosukhin}]{vaswani2017attention}
	Vaswani, A., Shazeer, N., Parmar, N., Uszkoreit, J., Jones, L., Gomez, A.~N.,
	Kaiser, {\L}., Polosukhin, I., 2017. Attention is all you need. In: Advances
	in neural information processing systems. pp. 5998--6008.
	
	\bibitem[{Wang et~al.(2018)Wang, Wang, Liu, Chen, Zhang, and
		Qi}]{wang2018towards}
	Wang, M., Wang, R., Liu, J., Chen, Y., Zhang, L., Qi, G., 2018. Towards empty
	answers in sparql: Approximating querying with rdf embedding. In:
	International Semantic Web Conference. Springer, pp. 513--529.
	
	\bibitem[{Wang et~al.(2017)Wang, Mao, Wang, and Guo}]{wang2017knowledge}
	Wang, Q., Mao, Z., Wang, B., Guo, L., 2017. Knowledge graph embedding: A survey
	of approaches and applications. IEEE Transactions on Knowledge and Data
	Engineering 29~(12), 2724--2743.
	
	\bibitem[{Wang et~al.(2014)Wang, Zhang, Feng, and Chen}]{wang2014knowledge}
	Wang, Z., Zhang, J., Feng, J., Chen, Z., 2014. Knowledge graph embedding by
	translating on hyperplanes. In: AAAI. Vol.~14. pp. 1112--1119.
	
	\bibitem[{Yan et~al.(2019)Yan, Janowicz, Mai, and Zhu}]{yan2019spatially}
	Yan, B., Janowicz, K., Mai, G., Zhu, R., 2019. A spatially explicit
	reinforcement learning model for geographic knowledge graph summarization.
	Transactions in GIS 23~(3), 620--640.
	
	\bibitem[{Yang et~al.(2015)Yang, Yih, He, Gao, and Deng}]{yang2014embedding}
	Yang, B., Yih, W.-t., He, X., Gao, J., Deng, L., 2015. Embedding entities and
	relations for learning and inference in knowledge bases. In: The Third
	International Conference on Learning Representations.
	
	\bibitem[{Zaheer et~al.(2017)Zaheer, Kottur, Ravanbakhsh, Poczos,
		Salakhutdinov, and Smola}]{zaheer2017deep}
	Zaheer, M., Kottur, S., Ravanbakhsh, S., Poczos, B., Salakhutdinov, R.~R.,
	Smola, A.~J., 2017. Deep sets. In: Advances in neural information processing
	systems. pp. 3391--3401.
	
	\bibitem[{Zhu et~al.(2016)Zhu, Hu, Janowicz, and McKenzie}]{zhu2016spatial}
	Zhu, R., Hu, Y., Janowicz, K., McKenzie, G., 2016. Spatial signatures for
	geographic feature types: Examining gazetteer ontologies using spatial
	statistics. Transactions in GIS 20~(3), 333--355.
	
\end{thebibliography}

\end{document}